\documentclass[preprint,12pt]{elsarticle}

\usepackage{graphicx}

\usepackage{amssymb}

\usepackage{algorithmic}
\usepackage{algorithm}

\def\be{\begin{equation}}
\def\ee{\end{equation}}
\def\ba{\begin{array}}
\def\ea{\end{array}}
\def\bea{\begin{eqnarray}}
\def\eea{\end{eqnarray}}
\def\beas{\begin{eqnarray*}}
\def\eeas{\end{eqnarray*}}

\newcommand{\bfc}{\mbox{\boldmath $c$}}

\newcommand{\bfu}{\mbox{\boldmath $u$}}

\newcommand{\bfw}{\mbox{\boldmath $w$}}

\newcommand{\bfA}{\mbox{\boldmath $A$}}

\newcommand{\bfJ}{\mbox{\boldmath $J$}}

\newcommand{\bfT}{\mbox{\boldmath $T$}}

\newcommand{\bfxi}{\mbox{\boldmath $\xi$}}

\newcommand{\bfphi}{\mbox{\boldmath $\phi$}}

\newcommand{\bfpsi}{\mbox{\boldmath $\psi$}}

\journal{Computers and Mathematics with Applications}
\begin{document}

\begin{frontmatter}



\title{Numerical integration on GPUs \\ for higher order finite elements}

\author[laddr2]{Krzysztof Bana\'{s}\corref{cor1}}
\ead{pobanas@cyf-kr.edu.pl}
\author[laddr2]{Przemys{\l}aw P{\l}aszewski}
\ead{pplaszew@agh.edu.pl}
\author[laddr1]{Pawe{\l} Macio{\l}}
\ead{pmaciol@pk.edu.pl}

\address[laddr2]{Department of Applied Computer Science and Modelling, \\
AGH University of Science and Technology, Mickiewicza 30, 
30-059 Krak\'{o}w, Poland}
\address[laddr1]{Institute of Computer Science, \\
Cracow University of Technology, Warszawska 24, 31-155 Krak\'{o}w, Poland}

\cortext[cor1]{Corresponding author}

\begin{abstract}
The paper considers the problem of implementation on graphics processors of numerical integration routines for higher order finite element approximations. The design of suitable GPU kernels is investigated in the context of general purpose integration procedures, as well as particular example applications. The most important characteristic of the problem investigated is the large variation of required processor and memory resources associated with different degrees of approximating polynomials. The questions that we try to answer are whether it is possible to design a single integration kernel for different GPUs and different orders of approximation and what performance can be expected in such a case.
\end{abstract}


\end{frontmatter}


\section{Introduction}
\label{}

In recent years graphics processors (GPUs) have gained widespread acceptance as an architecture useful for scientific computing. In many application domains their use brings manyfold increase in performance \cite{sc_gpu_review,Owens_2007}. Most often GPUs are used for computational kernels, the most computationally intensive parts of applications. 

In the finite element method (FEM) domain, the first cited case of using GPUs is \cite{Wu_2004} where finite elements are applied in interactive visualisation. Other examples include implementation for GPU clusters of higher order FEM approximations in earthquake modelling and wave propagation problems \cite{komatitsch09} or GPU implementations of some variants of discontinuous Galerkin approximation \cite{Klockner_2009}.

The finite element solution process can be divided into two parts: creation of the matrix for the system of linear equations, based on a suitable weak statement of the approximated problem and the solution of the system of equations (or some update of the vector of global degrees of freedom, based on the created matrix -- or several matrices like e.g. for certain time stepping schemes -- and the right hand side vector). 

The two parts can be implemented as different software components. The second component, the solver of linear equations, whose implementation can often neglect numerical and computational details associated with finite element approximation, is more frequently investigated in the context of high performance computing and especially GPU acceleration (see e.g. \cite{goddeke_solid,GoStMo_07scalability}). Sometimes the main stress is put on computational kernels of linear solvers, particularly linear algebra operations, that are properly optimized \cite{Volkov_2008,fujimoto_nvidia}. 

As a justification for such an approach, one can use the fact that quite often the time required for the solution of the system of linear equations strongly dominates the time for the whole solution process. This always happens in situations where weak statements employed are simple, degree of approximation is low and, moreover, the system of linear equations is ill-conditioned or non-definite, causing iterative solvers to converge slowly or  enforcing the use of direct solvers.

However for complex weak formulations and high orders of approximation the time for creation of system of linear equations can be higher than the time for its solution. The most time consuming part of the system creation is numerical integration of terms from the weak statement. Because of that, optimization of numerical integration routines has been for many years the subject of investigations in the higher order and spectral approximation communities \cite{dubiner,sherwin,schwab_integr,leszek_hp_book_v2,Vos_2010}.

Numerical integration in finite element codes is used to create entries for element stiffness matrices that are further assembled into the global stiffness matrix, the matrix for the associated system of linear equations. 
Again this two processes can be investigated together or treated separately.

The problem of combined integration and assembly for particular applications ported to GPUs has been described e.g. in \cite{Filipovic_assembly} for nonlinear elasticity and in \cite{Dziekonski_generation}
for electromagnetics. 
An extensive study presented in \cite{Klockner_2009} concerns 3D discontinuous Galerkin approximations for hyperbolic problems. Higher order tetrahedral elements are considered, with the special focus on particular operators associated with explicit time stepping mechanism and DG discretization. 

The approach presented in \cite{Klockner_2009} for DG problems follows a generally applicable technique that considers numerical integration and global assembly as steps of evaluation of FEM operators \cite{Kirby_FEM_operators}. This technique is important not only in the context of explicit time stepping methods, but also e.g. for the implementation of iterative solvers of linear equations and the application of boundary conditions from FEM weak statements. This approach has formed a basis for the attempts to create tools for automated generation of finite element codes \cite{Kirby_compiler,Logg_DOLFIN,Fenics_book}.

The continuation of these investigations in the context of GPU computing was recently described in \cite{Markall_2010,Markall_2013} where different assembly techniques for finite element approximations are considered. The performance of presented techniques is tested for 2D low order approximations. OpenCL and CUDA programming environments are compared and portability of OpenCL kernels between GPUs and CPUs is examined.

In the similar context (automated code generation and low order approximations) the problem of finite element integration is presented in \cite{Knepley_2013}.
The implementation is tested for 2D and 3D elliptic problems (Laplace and elasticity operators) and the performance obtained for GPUs is reported.

The three mentioned above papers suggest automatic generation of optimised code for each processor architecture, based on some abstract specification of computations, as a solution to an important implementation problem of the portability of software in face of diversity of multi-core architectures. The authors confirm the observations from other application domains \cite{Rul_10}, that also in the context of finite element assembly, even in the case of the same programming model (e.g. OpenCL) employed for different architectures, in order to obtain the best performance, it is necessary to produce separate versions of procedures for each architecture. 

Thorough investigations of the finite element assembly process on GPUs were presented in \cite{Cecka_2011}. In the article several methods of assembly, that imply also numerical integration techniques, were shown. The performance of the proposed techniques was tested for scalar problems and non-adaptive 2D meshes. The approach presented in \cite{Cecka_2011} was extended in \cite{Cecka_gems} to 3D problems, however for first order polynomials only. Nevertheless, the techniques presented in the papers can form the basis for various implementation strategies of the assembly process, also in the case where numerical integration is performed separately and different types of approximation are considered.

There may be several reasons for considering numerical integration separately from the global assembly. First, the solver employed for the solution of the system of linear equations may perform itself the conversion of element matrices to some specific sparse format of the global matrix \cite{iccs_hp_solver_2013}. Second, the assembly may be performed by a different hardware component (such an approach for large scale problems has been adopted e.g. in \cite{Dziekonski_assembly}). And finally, considering numerical integration independently of the global assembly gives software developers more flexibility in designing the final solution strategy.

As an additional argument one may consider also the relative importance of both operations. The assembly of an element matrix to the global stiffness matrix involves three memory references and one floating point operation. For higher order approximations that we consider in the paper, computing a single entry in the element matrix may require, as we show in detail later in the paper, hundreds of memory accesses and floating point operations. Even considering that for numerical integration the references can be realized using faster local memories without data races, still the proper implementation of numerical integration can be considered as having its own merits and importance, independent of the global assembly.

When porting computations to graphics processors, one encounters another specific problem -- the issue of precision. Some GPUs do not support double precision arguments, some perform double precision operations much slower than single precision. 

It has been proven in many studies that the solution of linear equations, especially for large scale problems, requires the use of double precision unknowns, because of accumulation of round-off errors. When investigating the implementation of that stage of finite element solution process, the solution might be to select the phase of calculations or particular operations that can be performed in single precision and after the single precison phase perform improvement of results \cite{goddeke_mixed,Dongarra_mixed}. 

Sometimes, when the solution process does not require the solution of the system of linear equations, due to the requirements of a particular application, the whole finite element simulation process can be performed in single precision. This can happen for example in the case of explicit algorithms for time dependent problems \cite{Komatitsch_2010}. 

Even in the case of performing the stage of linear system solution, numerical integration can be performed in single precision. It forms a part of the whole approximation process and has to satisfy specific requirements concerning accuracy, in order to guarantee the convergence of finite element solutions \cite{ciarlet}. For many problems, due to the approximate character of numerical integration itself, performing calculations in single precision does not introduce errors large enough to be comparable to discretization errors \cite{Cecka_2011}. In such cases numerical integration can be performed in single precision and the resulting entries to element stiffness matrices or the global stiffness matrix can be promoted to double precision before the solution of the system of equations.

For some problems, e.g. for problems with terms of different orders that scale in a different way, double precision is required also for numerical integration \cite{leszek_hp_book_v1}. However, this can be considered a special case, while general purpose calculations should be performed in single precision, due to shorter execution times and lower memory requirements. The advantages of single precision computations can be especially pronounced for higher order calculations. 

\subsection{Current contribution}
We consider the problem of numerical integration for higher order 3D finite element approximations. The main feature of the problem that we address is the large variation of required resources, processing time and memory, for different orders of approximation, i.e. different degrees of polynomials employed as finite element shape functions. 

We investigate the characteristics of the problem and try to design a portable kernel for different problems and different processor architectures. Hence, we restrict ourselves to OpenCL programming model. It allows us to construct kernels for graphics processors supporting the model (currently NVIDIA, AMD and Intel hardware), as well as for CPUs. Our previous investigations \cite{imcsit_10} as well as results reported in other papers (e.g. \cite{Markall_2013}) show that OpenCL implementations achieve performance only several percent lower than native CUDA model for NVIDIA GPUs.

We situate our research in the context of creating a general purpose finite element framework ModFEM \cite{cmms_modfem_2013}, for continuous and discontinuous Galerkin approximations. The code has modular structure in which different software components are responsible for different phases of finite element calculations \cite{iccs04}. Each component can exist in several variants and final applications for different problems and solution strategies are created using suitable compilation options.

The code has also a layered structure with separate modules responsible for shared memory computations, that form a core layer and that are further extended to distributed memory environments by suitable overlays \cite{dd15,iccs06}, belonging to a different layer. The code operates on unstructured adaptive meshes with proper load balancing for parallel execution \cite{ppam03}. The code has abilities to run on clusters with message passing and is currently ported to different new computer architectures, such as GPUs and heterogeneous processors, like CBE \cite{ppam_09_cell}.

In each particular code created by ModFEM framework there is a separate approximation module that is responsible for the creation of element stiffness matrices. The matrices are further transferred to linear equations solvers using a generic interface \cite{cvs}. Thanks to the layered structure of the code there is only one numerical integration routine, in the core layer, for both, shared and distributed memory environments. The interface between numerical integration procedure and the solver of linear equations allows for employing different, direct and iterative, solvers of linear equations, possibly using thin adapters \cite{cames_hp_solver}.

The same integration routine is also used for different types of meshes and problems solved. This approach proved successful in many application domains \cite{jcp1,bea_compgeo2,cmms_welding_2013}. As a main problem of the current investigations we pose the problem whether it possible to maintain this portability and genericity by creating a single kernel executed on GPUs for different problems, meshes and computer architectures.

We consider numerical integration only, leaving the problem of assembly (and also application of boundary conditions) to different software and possibly hardware components. In our view this smaller grain of considered GPU code gives more flexibility in designing solution strategies, where CPU and GPU cores can work together, each group performing the tasks it is most suitable for \cite{Dziekonski_generation}.
In order to maximize this flexibility of the approach we consider different options for passing the results of calculations. The algorithms presented in the paper do not prevent the use of different assembly techniques, such as e.g. presented in the discussed above studies devoted to the subject.

We attempt to investigate the possibility of creating a general purpose numerical integration procedure executed on graphics processors that can either replace or complement the CPU integration routine. We do not attempt to create GPU-only code. Considering relatively small sizes of memory for current accelerators with GPUs, in the order of several GBytes, this would limit the sizes of problems that we solve. Since we aim at large scale problems we want to use large memories, in the range of hundreds of GBytes, for each shared memory node. 

The use of GPUs as accelerators for CPUs, corresponds also to our distributed memory strategy where we assume two separate levels of decomposition. GPU calculations fit into shared memory level calculations controlled by CPU code and due to the design philosophy can be directly used in distributed memory calculations, based on CPU code and its message passing overlays.


Following the arguments presented above, we assume that our general purpose finite element numerical integration procedure does not require double precision calculations and we restrict our interest to single precision implementations. We plan to consider extensions to double precision for particular application domains that would pose such requirements.

The investigations presented in the paper form the continuation of an earlier work. In \cite{ppam_09_gpu} we presented numerical integration on GPUs for 2D linear elasticity, CUDA environment and older generations of GPUs. This work was extended to OpenCL programming model in \cite{imcsit_10} and to 3D Laplace operators in \cite{iccs_10_gpu}. We investigate separately numerical integration procedures for CBE processors in \cite{ppam_09_gpu} and \cite{cmwa_12_ni_CBE}.

The paper is organised in the following way:
\begin{itemize}
\item Section \ref{problem_formulation}
defines the problem of numerical integration for finite element approximations
\item Section \ref{programming_model}
describes the programming model that we use and its mapping to selected GPU architectures
\item Section \ref{parallel_implementation}
contains details of several versions of numerical integration procedures for GPUs that we developed
\item Section \ref{computational_tests}
presents results of tests performed to assess the performance of the code
\item Section \ref{conclusions}
contains final remarks.
\end{itemize}

\section{Problem formulation}
\label{problem_formulation}
In the current section we define the problem of numerical integration for which we will further design GPU kernels. We restrict ourselves to defining the notation that we use and computational aspects of the problem. For more information on numerical aspects of integration we refer to the works presented in the introduction and for the whole context of higher order finite element approximation to any of the excellent textbooks on the subject (e.g. \cite{leszek_hp_book_v1}, \cite{solin}).

3D finite element numerical integration and assembly consist of a loop over elements and faces and possibly some other mesh entities in order to create entries to the global stiffness matrix and the global right hand side vector. The creation can involve different operations -- integration over elements, integration over faces and other forms of application of boundary conditions, application of other constraints, etc. The most time consuming part, especially for higher orders of approximation, is the calculation of element integrals, and following the discussion presented in the introduction, we restrict our attention to this process.


For numerical integration we use quadratures with the order determined by the order of approximation and the characteristics of the problem solved. Integration is always performed on a reference element, following the suitable change of integration domain from a real element in physical space to the reference element. Within a 3D reference element quadrature points have coordinates $\bfxi^Q$:
\[
\bfxi^Q[i_Q][i_S];\; \; i_Q = 1, 2, ..., N_Q; \;  i_S = 1, 2, 3;
\]
and weights $\bfw^Q$
\[\bfw^Q[i_Q], i_Q = 1, 2, ..., N_Q\]
 with $N_Q$ denoting the number of quadrature points.

Computed integrals correspond to different terms in the weak statement of the problem considered. The problem dependent contribution consist in defining a set of coefficients for numerical integration. There is a different coefficient for each combination of indices, that we denote by $i_D$ and $j_D$, corresponding to different spatial derivatives for test and trial functions  respectively, including terms with functions itself, that we associate with $i_D=0$ and $j_D=0$.  Moreover, for vector problems, the same approximation may be used for different vector components (different unknowns in the solved system of PDEs) and hence for each combination of indices $i_D$  and $j_D$ there may be a small matrix of coefficients with the dimension $N_{E}$ equal to the number of equations in the solved system of PDEs. In such a case the array of coefficients may be defined as:
\[
\bfc[i_{E}][j_{E}][i_D][j_D];\; \;  i_{E},j_{E} = 1, 2, ..., N_{E}; \; i_D,j_D = 0,1,2,3; \;
\]

In the most general, non-linear or quasi-linear case there may be different values of coefficients at each integration point. 
Hence for the generic numerical integration algorithm we have to consider an array of coefficients of the form:
\[
\bfc^Q[i_Q][i_{E}][j_{E}][i_D][j_D];\; \;  i_Q = 1, 2, ..., N_Q; i_{E},j_{E} = 1, 2, ..., N_{E}; i_D,j_D = 0,1,2,3 
\]

As it was stated above, the integration of terms from the finite element weak statement is performed using the change of variables, from a real element to a reference element, of one of several possible types (e.g. tetrahedron, prism, cube, to name the most popular). Hence the derivatives with respect to physical coordinates, that appear in the weak statement, have to be computed by the chain rule, using derivatives of trial and test functions with respect to reference coordinates. These calculations involve the Jacobian matrix of the transformation from the real element to the reference element. The matrix is usually obtained by inverting the matrix of its inverse transformation, the transformation from the reference to the real element. This last transformation can be easily obtained from the description of geometry of real elements. The Jacobian matrix of the transformation    from the reference to the real element is also used for computing the volume element in integrals, and later in the paper when the notion of the Jacobian matrix is used it denotes this matrix. 

The transformation from the reference to the real element is performed using the representation of the geometry of the real element. Different options are possible here, with the simplest of defining linear or multi-linear elements, whose geometry is determined by the position of their vertices. More complex cases include popular isoparametric elements, where the geometry is specified using polynomials of the same order as the solution \cite{solin} and recently introduced isogeometric analysis, where elements are geometrically represented using spline approximations \cite{hughes_isogeometric_book}.

Element stiffness matrices are computed for finite element approximations to trial and test functions, represented as linear combinations of finite element basis functions, that in turn are constructed using element shape functions. The number of shape functions for an element, $N_{\mathrm{sh}}$  (equal to the number of local degrees of freedom for the element) is related to the order of approximation and possible particular choices of approximation. 
In order to present concisely the numerical integration problem we denote shape functions and their derivatives by a single array 
\[
\bfphi[i_D][i_{DOF}];\; \;  i_D = 0,1,2,3; \; i_{DOF} = 1, 2, ..., N_{\mathrm{sh}}
\]
where for the first index, its value 0 refers to shape functions and the values 1,2,3 refer to derivatives with respect to local, reference element coordinates, $\bfxi = \{ \xi_1, \xi_2, \xi_3 \} $, and the second index corresponds to different degrees of freedom.

In order to compute integrals we use the values of shape functions and their derivatives at all integration points:
\[
\bfphi^Q[i_Q][i_D][i_{DOF}]; \; \; i_Q = 1, 2, ..., N_Q; \; i_D = 0,1,2,3; \; i_{DOF} = 1, 2, ..., N_{\mathrm{sh}};  
\]

These values are the same for all real elements of a given approximation order. In calculations of terms from the weak statement we use the values of shape functions and their derivatives with respect to physical coordinates. These values will be denoted by:
\[
\bfpsi^Q[i_Q][i_D][i_{DOF}];\; \; i_Q = 1, 2, ..., N_Q; \; i_D = 0,1,2,3;\;  i_{DOF} = 1, 2, ..., N_{\mathrm{sh}};  
\]

The derivatives of shape functions with respect to physical coordinates are different for each element, since they are computed using the entries of the Jacobian matrix of the transformation from the real element to the reference element, that are different for each element. The Jacobian matrix of the transformation from the real element to the reference element is also used for computing derivatives of the solution at previous non-linear iterations and/or previous time steps. This derivatives (and values of the solution itself) are used in non-linear and time dependent problems to compute problem dependent coefficients $\bfc$.

The finite element numerical integration problem for a single element can be defined in the following way:
\vspace{2mm}

Given a set of data defining the geometry of an element, a set of degrees of freedom corresponding to solutions at previous non-linear iterations and/or previous time steps and a set of data specifying problem dependent input to the integration procedure create an element stiffness matrix
\[
\bfA^e[i_{E}][j_{E}][i_{DOF}][j_{DOF}];\; \;  i_{E},j_{E} = 1, 2, ..., N_{E};\;  i_{DOF},j_{DOF} = 1, 2, ..., N_{\mathrm{sh}};
\]
corresponding to the finite element formulation of the problem solved.
\vspace{2mm}

\begin{algorithm}
\caption{Basic numerical integration algorithm for a single element}
\label{num_int_simple}
\begin{algorithmic}[1]
\STATE read problem dependent coefficients, $\bfc$
\STATE read geometry data for element
\STATE possibly read ''old'' element degrees of freedom from previous iterations/time steps
\STATE initialize element stiffness matrix, $\bfA^e$
\STATE prepare quadrature data, $\bfxi^Q$ and $\bfw^Q$
\FOR{$i_Q=1$ \TO $N_Q$} 
\STATE read values of shape functions and their derivatives with respect to local element coordinates, $\bfphi^Q[i_Q]$
\STATE read or calculate Jacobian matrix, its determinant ({\bf det}) and its inverse
\STATE calculate derivatives of shape functions with respect to physical coordinates, $\bfpsi^Q[i_Q]$
\STATE based on $\bfc[i_Q]$ and ''old'' degrees of freedom calculate coefficients at quadrature point, $\bfc^Q[i_Q]$
\FOR{$i_{DOF}=1$ \TO $N_{\mathrm{sh}}$}
\FOR{$j_{DOF}=1$ \TO $N_{\mathrm{sh}}$}
\FOR{$i_D=1$ \TO $N_D$}
\FOR{$j_D=1$ \TO $N_D$}
\FOR{$i_E=1$ \TO $N_E$}
\FOR{$j_E=1$ \TO $N_E$}
\STATE $\bfA^e[i_{E}][j_{E}][i_{DOF}][j_{DOF}]+=$
\STATE $\; \; \; \; \; \; \; \; ${\bf det}$\times \bfw^Q [i_Q] \times \bfc^Q[i_{E}][j_{E}][i_D][j_D][i_Q] \times$
\STATE $\; \; \; \; \; \; \; \; \bfpsi^Q[i_D][i_{DOF}][i_Q] \times \bfpsi^Q[j_D][j_{DOF}][i_Q]$
\ENDFOR
\ENDFOR
\ENDFOR
\ENDFOR
\ENDFOR
\ENDFOR
\ENDFOR
\end{algorithmic}
\end{algorithm}

We conclude the definition of the problem by providing Algorithm \ref{num_int_simple}, a generic procedure that calculates the entries to the element stiffness matrix. The procedure corresponds to mathematical expressions involved in numerical integration, indicating necessary summations. 

From the mathematical point of view, the order of performing the three outermost loops of Algorithm \ref{num_int_simple} is irrelevant (but implies some changes in details of the algorithm). In our implementation we follow the common practice from classical finite element codes, corresponding directly to Algorithm \ref{num_int_simple}, where the loop over integration points is the outermost loop. Thanks to this we do not have to store in some fast memory the values of all shape functions at all integration points, instead we need only quickly accessible storage for the values of shape functions and their derivatives at a single integration point. 

The other option, with precomputed values of all shape functions at all integration points and the loops over degrees of freedom (shape functions) as the outermost loops, is also possible, especially for low order 2D approximations \cite{Cecka_2011,Markall_2013}.

In practice, calculations are often performed in a way that is different that specified by Algorithm \ref{num_int_simple}. The four innermost loops are small, with the ranges for $i_D$ and $j_D$ equal 4 and usually less or equal 5 for $i_E$ (3 for elasticity, 4 for incompressible Navier-Stokes, 5 for compressible Navier-Stokes). This suggests to manually unroll all the loops.

Moreover arrays $\bfc$ are usually sparse and substantial savings can be obtained, when instead of performing the four innermost loops from Algorithm \ref{num_int_simple}, suitable, optimized calculations are performed for a block of entries of the matrix $\bfA^e$ associated with a single pair  $i_{DOF}, j_{DOF}$. This leads to Algorithm \ref{num_int_real}, that we further implement for graphics processors.

\begin{algorithm}
\caption{Practically applicable numerical integration algorithm for a single element}
\label{num_int_real}
\begin{algorithmic}[1]
\STATE read problem dependent coefficients, $\bfc$
\STATE read geometry data for element
\STATE possibly read ''old'' element degrees of freedom from previous iterations/time steps
\STATE initialize element stiffness matrix, $\bfA^e$
\STATE prepare quadrature data, $\bfxi^Q$ and $\bfw^Q$
\FOR{$i_Q=1$ \TO $N_Q$} 
\STATE read or calculate values of shape functions and their derivatives with respect to local element coordinates, $\bfphi^Q[i_Q]$
\STATE read or calculate Jacobian matrix, its determinant ({\bf det}) and its inverse
\STATE calculate derivatives of shape functions with respect to physical coordinates, $\bfpsi^Q[i_Q]$
\STATE based on $\bfc[i_Q]$ and ''old'' degrees of freedom calculate coefficients at quadrature point, $\bfc^Q[i_Q]$
\FOR{$i_{DOF}=1$ \TO $N_{\mathrm{sh}}$}
\FOR{$j_{DOF}=1$ \TO $N_{\mathrm{sh}}$}
\STATE update a block of entries of $\bfA^e$ associated with a pair $i_{DOF}, j_{DOF}$
in a manner implied by the structure of non-zero entries of the array of coefficients $\bfc$, multiplying 
each non-zero entry of $\bfc^Q$ associated with a pair $i_{D}, j_{D}$ by
\STATE $\; \; \; \; \; \; \; \; ${\bf det}$\times \bfw [i_Q] \times  \bfpsi^Q[i_D][i_{DOF}][i_Q] \times \bfpsi^Q[j_D][j_{DOF}][i_Q]$
\STATE and performing suitable summations over ranges of $i_D$ and $j_D$
\ENDFOR
\ENDFOR
\ENDFOR
\end{algorithmic}
\end{algorithm}


To assess the particular character of the problem of numerical integration for 3D approximations with different orders of approximation $p$, we present in Table \ref{tab_1} the sizes of the arrays involved in calculations for an example 3D reference element --- the prism with shape functions being products of polynomials from a set of complete polynomials of a given order for triangular bases and 1D polynomials associated with vertical direction. As an example quadrature, the most popular in finite element codes, Gaussian quadrature is selected.
As can be seen from the table, the number of times the innermost calculations are performed for $p=7$ equals 27,869,184. That is one of reasons why for higher orders of approximation special techniques for integration are designed \cite{schwab_integr,Vos_2010}. 

The order from which it becomes advantageous to switch to different techniques that presented in the current paper depends on the characteristics of the problem statement and the computing environment. The value of 7 that we choose for our study is to certain extent arbitrary (in some cases it can be profitable to switch to different techniques of integration even from orders in the range 4-5). Nevertheless, for the rest of the paper we consider  for our higher order finite element approximations the range of polynomial degrees $p$ from 2 to 7. 

\begin{table}
\begin{center}
\begin{tabular}{|l|r|r|r|r|r|r|r|}
\hline
&  \multicolumn{7}{|c|}{Degree of approximation p}  \\
  & 1 & 2 & 3 & 4 & 5 & 6 & 7  \\
\hline
$N_{\mathrm{sh}}$
&6&	18&	40&	75&	126&	196&	288\\
\hline
$N_Q$
&6&	18&	48&	80&	150&	231&	336 \\
\hline
\end{tabular}
\end{center}
\caption{Parameters determining the computational characteristics of
 3D finite element numerical integration -- the number of shape
  functions and the
  number of Gaussian integration points for the standard prismatic element and different degrees
  of approximating polynomials}
\label{tab_1}
\end{table}

The second important set of parameters determining the character of numerical integration algorithm, implied by the data from Table \ref{tab_1} and especially important for GPU computations, are the numbers of entries in arrays involved in calculations, presented in Table \ref{tab_2}.
The first row of data in Table \ref{tab_2} presents the
size  of  array containing shape functions and their derivatives used for computing integrals
for a single integration point, while the last row shows the 
respective sizes
for all integration points. 
The large variation of the sizes presented in Table \ref{tab_2} enforces a special design of integration routines, when aiming at a generic code for all approximation orders.

\begin{table}
\begin{center}
\begin{tabular}{|l|r|r|r|r|r|r|r|}
\hline
The number of &  \multicolumn{7}{|c|}{Degree of approximation p}  \\
entries (blocks) in: & 1 & 2 & 3 & 4 & 5 & 6 & 7  \\
\hline
$\bfxi^Q$ and $\bfw^Q$
& 24 &	72 & 192 &	320 &	600 &	924 &	1344 \\
\hline
$\bfphi^Q[i_Q]$
& 24 &	72 &	160 &	300 &	504 &	784 &	1152\\
\hline
$\bfphi^Q$
& 144 &	1296 &	7680 &	24000 &	75600 &	181104 &	387072 \\
\hline
$\bfA^e$ (blocks)
& 36 & 324 &	1600 &	5625 & 15876 & 38416 & 82944 \\
\hline
\end{tabular}
\end{center}
\caption{The number of entries (blocks) in arrays used in 3D numerical integration
algorithm for the standard prismatic element and different orders of approximation: quadrature points and weights, values of shape functions and their derivatives at single integration point,  values of shape functions and their derivatives at all integration points, $i_E \times i_E$ blocks of element stiffness matrix }
\label{tab_2}
\end{table}

\section{Programming model for GPU implementation of numerical integration and its mapping to GPU architectures}
\label{programming_model}

In the current section we briefly present the programming model that we use in our parallel numerical integration algorithms for GPUs. We describe also basic performance optimization techniques specific to GPU computing, that we employ in our codes.

\subsection{Programming models for GPUs}

CUDA \cite{cuda_guide}  and OpenCL \cite{OpenCl,gaster2011opencl} are the two most popular programming models for GPUs. CUDA was the first platform for programming GPUs that gained widespread acceptance and popularity. Still it is the most advanced programming environment, aimed at getting the most of graphics hardware produced by NVIDIA.

The OpenCL programming model and environment was created later, however, from the beginning it targeted a broader spectrum of hardware, including, apart from GPUs, standard multi-core CPUs and hybrid architectures, like Cell Broadband Engine (CBE). In the context of creating a general purpose, portable code for finite element simulations we concentrate on the OpenCL programming model.

Despite different targeted architectures, CUDA and OpenCL are very similar. Most of CUDA codes, that do not use advanced capabilities of the computing environment, can be easily transferred to OpenCL. Most of notions used in one programming model have direct analogues in the other.
Software design presented in a sufficiently generic manner for one of the models can be considered as a design for both. We believe that the procedures that we design for OpenCL can be directly ported to CUDA, with no performance loss. They can be further optimized for NVIDIA hardware using techniques specific to CUDA.

In order to make our presentation easily accessible to persons more accustomed to CUDA programming, when introducing OpenCL notions, we recall in parantheses respective CUDA names if they differ from the ones used in OpenCL. In the description of GPU algorithms we use a mixture of the two styles, that we consider intuitively clear.

\subsection{OpenCL programming model}

OpenCL code is compiled and run on a given platform, that represents the environment for code execution. Each platform provides a set of devices (CPUs, GPUs, accelerators). For each platform there exists a host system that runs standard code and manages the execution of a set of kernels on devices. Kernels are functions written in the OpenCL language, a slightly modified version of the C language. Since the specification does not provide separate notions, by kernels we will understand both, functions in the form of source code, as well as in binary form that is send for actual execution.  

Devices are composed of compute units, which are further divided into processing elements. Individual threads of execution (called work-items in OpenCL specification -- the notion that we will not use in the paper) are running on processing elements, that are in principle capable of executing scalar and vector instructions. 

For running on compute units, threads are grouped into work-groups (thread-blocks in CUDA nomenclature), with one work-group assigned to a single compute unite (each compute unit can, however, have many work-groups assigned to it). Threads within a single work-group execute concurrently, can share some data in fast memory and can be synchronised using fast system calls. Different work-groups are scheduled independently.

For running on a device, thread work-groups are further combined into the whole set of threads executing a kernel. Threads within a work-group have local identifiers, that may have form of 1, 2 or 3 dimensional arrays. Global identifiers for threads within the whole executing set can have similar form. We use simple one dimensional local and global numbering of threads. Work-groups have also their indices, and, in the same way as in the case of threads, we use one dimensional numbering of work-groups.

OpenCL execution model specifies the events that has to occur in order for device code to be run. 
The first phase includes initialization of the specific OpenCL data structures and checking the available devices. Then the code to be executed on the devices has to be read from source (a distinguishing feature of OpenCL) or binary file, possibly compiled and prepared in the form of a kernel. During compilation several options can be passed to the OpenCL compiler.

Kernels can accept arguments, hence, before executing a kernel, the space for arguments has to be allocated on device and arguments send to the device memory. The host code can allocate space for variables and arrays in different explicitly available to programmers types of memory.

The memory transactions are performed using a typical for OpenCL strategy, by sending a request to the OpenCL management layer. The requests are then realized asynchronously to the host code. In the same way the requests for kernel execution and transferring back data from device memory to the host memory are realized.

OpenCL memory model defines several types of memory regions explicitly available to programmers. Different memory objects can be created for OpenCL kernels with different mappings to hardware resources. We briefly describe simple classification of variables that we use in our calculations.

Individual variables defined in kernels belong to private (local in the CUDA nomenclature) memory. Each thread has its own copy of each variable, preferably stored in registers. 

Assignment to other memory regions is achieved through specific qualifiers. We use the arrays stored in the following types of memory:
\begin{itemize}
\item
global -- such arrays are visible to all threads executing the kernel
\item
constant -- being a part of global memory with read only access for threads
\item
local -- arrays in this fast memory are shared by threads in a single work-group 
\end{itemize}

Since the notion of local memory has different meanings in OpenCL and CUDA terminology, we do not use it in the rest of the paper. Instead, for OpenCL local memory we use the CUDA term ''shared memory'' that in our view better reflects its character. 

Since OpenCL is aimed at portability of created codes, it contains procedures that allows for adapting to different platforms and devices. The code can query the environment to get information on many available resources. We employ several of such functions, the ones that we use will be described in the section devoted to parallel GPU implementation of numerical integration.

\subsection{Mapping of OpenCL and CUDA notions to GPU hardware resources}
\label{soft_to_hard}
In this section we briefly describe the hardware resources available to CUDA and OpenCL programmers. Since the GPU hardware changes rapidly, we do not give technical details, that are different for different architectures and different generations of GPUs. We try to provide a generic description that should serve as a justification for several of design choices that we make. We restrict the description to NVIDIA and AMD GPUs as the two most popular families. 

Explicitly defined OpenCL and CUDA abstractions are mapped to different hardware resources for particular GPU architectures. Compute units are called streaming multiprocessors for NVIDIA hardware, processing elements are often called CUDA cores for NVIDIA and stream cores for AMD. The details of GPU organization can be complex and the mapping of compute units and processing elements to hardware units may differ for different architectures \cite{cuda_guide,AMD_APP_guide}.

The memory hierarchy is implemented in AMD and NVIDIA architectures in a similar way. There are registers, off-chip DRAM for global and constant memory and on-chip hardware for shared and cache memory (constant memory is cached, the global memory is cached in recent generations of GPUs). The access times and bandwidths for different memory regions vary for different architectures. The variations include also relative speeds of different memory types. Specifically, shared memory may be almost as fast as registers for certain architectures (e.g. NVIDIA Fermi) or several times slower (e.g. AMD Evergreen). 

One important thing to mention, is handling of automatic variables, private to each thread. They are preferably stored in registers, but if too many registers are required by a kernel, either the kernel is not executed or register spilling occurs. For some architectures, spilled variables are stored in global memory, that significantly slows down the execution of a kernel. Recent GPUs utilize cache memory for variables spilled from registers, that makes the problem of minimizing the number of registers in a kernel less crucial. 

We use the term GPU memory for the global, device memory of GPUs. For each architecture that we consider, GPU memory is separate from the host memory and connected to it through PCIe bus. Slow PCIe connection should be taken into account when designing GPU kernels, as well as the fact that memory initialization forms another source of overhead for GPU execution.

The most important fact related to mapping software to hardware and not explicitly visible to programmers is the way threads are scheduled for execution. Both NVIDIA and AMD use SIMD style, where threads are grouped into sets that are called warps for NVIDIA and wavefronts for AMD. All threads in such a warp/wavefront execute one instruction  at a time.

If the instruction executed by a warp/wavefront is a global memory access and the locations accessed by different threads do not form one of specific patterns, one memory access instruction is executed using many memory transactions. However the proper arrangement of accesses allows hardware to, as it is called in CUDA nomenclature, coalesce accesses to different locations into a single memory transaction. 

Still such a transaction takes many clock cycles. In order to prevent pipeline stalls several warps/wavefronts have to be executed concurrently on a compute unit. We try to ensure in our kernels that many warps/wavefronts are scheduled concurrently, as a key ingredient for achieving high performance on GPUs. 

The problem of proper arrangement of memory accesses concerns not only the global GPU memory, but also the shared memory of compute units, that is implemented in hardware using many memory banks that have to be accessed in a way that eliminates bank conflicts. One simple rule, that should ensure the proper memory usage, is to make threads access subsequent elements of arrays storing single precision floating point numbers. 

We adopt this rule in the design of our kernels. Hence, whenever we talk about accessing vectors and matrices in memory, either global or shared, we mean accessing 1D arrays with subsequent entries read or written by subsequent threads.

It can be seen from the description above that the size of warps/wavefronts is an important characteristic of hardware. The number of threads in a work-group should be a multiple of warp/wavefront size. However the details of execution for different GPUs make the picture less clear \cite{cuda_guide,AMD_APP_guide}. We adopt, as a rule of thumb, based on the descriptions of utilized architectures, that the size of work-groups in our kernels forms a multiple of 64 (the size of two warps for NVIDIA and one wavefront for recent AMD GPUs).

\subsection{Further performance optimizations for GPUs}
Apart from standard optimizations that we perform, such as removing dependencies, reducing the strength of expressions, etc. and the guidelines presented in the previous section, we utilize several further rules related to specific character of GPUs.

One rule, is to maximize the number of threads executing a kernel. This is done not only due to the fact that GPUs have many processing elements, but also the fact that in order to hide memory access latency the number of threads ready to be executed should be a multiple of the number of processing elements. By which factor however, it depends on particular GPU architecture. We assume that at least several threads per processing element should be designated.

This can be achieved in two ways. One is to increase the size of a single work-group. By doing this the number of threads for a single compute unit is also increased. The second way is to increase the number of work-groups. Different work-groups can be scheduled to the same or to different compute units, hence the total number of threads for a GPU grows.

Which strategy to choose depends on several factors. One of such factors is the amount of memory resources required by threads. Increasing the number of threads, increase the required number of registers. Increasing the number of work-groups increase the required amount of shared memory. This may suggest as a good compromise to increase the number of threads within one work-group, up to the optimal maximum for a given compute unit, and then increase the number of work-groups to spread as many threads as possible across many compute units. The strategy has to be accompanied by a proper kernel design with minimal possible register and shared memory usage.

As we mentioned already, in designing our kernels we aim primarily at portability of the code across different problems considered and different processor architectures. Hence, we do not describe here and do not use specific optimizations targeted at particular architectures \cite{harris_cuda,gaster2011opencl}. This leaves the space for further improvements, that can yield better execution performance in particular circumstances.

\section{Parallel implementation of numerical integration for GPUs}
\label{parallel_implementation}

As we already mentioned in the introduction, the most natural way of parallelizing numerical integration algorithm is to parallelize the loop over elements. When applied to GPUs, this approach implies that the number of elements processed at once by a GPU is equal to the number of threads. GPU has to hold the state of each thread, hence the number of threads executed simultaneously may be limited by insufficient GPU resources. 

We follow a different approach with two level parallelization. At a higher level, loop over elements is parallelized. Still however, single element calculations are further destined for multithreaded execution.

There are two options for parallelization of numerical integration for a single element. One is to parallelize the loop over integration points (line 6 in Algorithm \ref{num_int_real}) and the second is to parallelize the two loops over blocks of stiffness matrix (lines 11 and 12 in Algorithm \ref{num_int_real}).

The loop over integration points is harder to parallelize, due to inherent data race in updating entries in $\bfA^e$ with contributions from different integration points. Moreover, the degree of concurrency is lower than for the double loop over blocks of $\bfA^e$. Hence in our algorithm we decided to parallelize the double loop over blocks of element stiffness matrix. This strategy can be classified as data parallel approach, since it corresponds to suitably decomposing the matrix $\bfA^e$, assigning different parts to different threads and performing calculations following the "owner computes" rule.

When applied to classical multi-core architectures, with cores equipped with large caches of different levels, the parallelization can be achieved in a straightforward manner, e.g. by introducing several OpenMP \cite{Chapman_OpenMP_2007} directives. In the standard way, one can consider row-wise, column-wise or block-wise decompositions of $\bfA^e$. 

For massively multi-core architectures of GPUs the degree of concurrency in such cases may turn out to be insufficient and lead to low performance of execution. Therefore we consider another strategy. The whole element stiffness matrix is stored as a single vector and this vector is decomposed into as many parts as there are threads at this level of parallelization. 

The number of threads and other execution characteristics are calculated taking into account small resources available to threads, an important limitation for designing algorithms on GPUs. This limitation often lead to complex algorithms and data management strategies necessary to obtain satisfactory performance.

\subsection{Kernel design}
The design of our parallel GPU code for higher order numerical integration is based on several assumptions:
\begin{itemize}
\item
single kernel invocation is performed for a large set of finite elements -- this is necessary since kernel invocation incur substantial overhead that has to be amortised over as many kernel operations as possible 
\item
one element is assigned to one work-group of threads, however one work-group operates on a set of elements, forming a subset of all elements associated with kernel execution
\item
for a single work-group and a single element parallelization is performed in the data-parallel manner, with each thread performing calculations for a set of entries in the element stiffness matrix
\end{itemize}

Several further design decisions have to be made then. How many threads should be in a work-group? How many entries of $\bfA^e$ should a single thread process? How arrays of shape functions should be accessed by threads? All these questions should be answered based on characteristics of the performed calculations in the light of great variations of the sizes of data structures associated with different orders of approximation.

One of such decisions concerns the use of values of shape functions. For sequential codes and low order approximations it may happen that computing the values of shape functions ''in flight'' may present no overhead as compared to reading them from memory. However for GPU execution, the facts that computing shape functions and their derivatives has small degree of concurrency (some parts of calculations are inherently sequential) and that these values are the same for all elements processed by the kernel, make precomputing the values and storing them in arrays in memory the only solution that can lead to high performance. Since the sizes of such arrays, presented in Table \ref{tab_2}, exclude the use of some faster memory, they are stored in global GPU memory.

The size of the element stiffness matrix, at least for the higher considered orders of approximation, is also too big to fit into fast GPU memory and $\bfA^e$ has to be kept in the slower global memory. It would be inefficient to perform summation of contributions from different integration points in global memory, hence it is performed using temporary variables in fast memory. We consider two choices: the values are stored either in registers (individual for each thread) or in shared memory (common to a work-group). 

If threads are using registers for storing computed values of $\bfA^e$ than for a single loop over integration points as many blocks of $\bfA^e$ as there are threads in a work-group are computed. After summation, the values are written to global memory and in the next step the threads has to consider new entries to be summed up. However this step requires repeating calculations in lines from 6 to 9 in Algorithm \ref{num_int_real}, including shape function reads and calculations related to Jacobian matrices, many of which are redundant. The whole procedure reduces to introducing another loop in Algorithm \ref{num_int_real}. Before the loop over integration points, there must appear a loop over parts of stiffness matrix processed at once by all threads in a work-group. 

The number of parts into which the whole element stiffness matrix $\bfA^e$ is divided can be simply computed based on the stiffness matrix size and the number of threads in a work-group. The number of parts is equivalent to the number of redundant repetitions of shape function reads and calculations related to Jacobian matrices. This redundant calculations form an overhead as compared to sequential code and can lead to lower parallel speed-up. We investigate two strategies to make this overhead as small as possible.

The first strategy is simple and consists in making the calculations in lines 6-9 in Algorithm \ref{num_int_real} as fast as possible. This includes the option for reading, instead of computing, the necessary terms related to Jacobian matrix and its inverse (we will call them later Jacobian terms). Since the calculations are inherently sequential this single thread section significantly slows down kernel calculations. However, the option of sending data implies that the values of Jacobian terms for all elements processed by the kernel and for all integration points are computed by the host code and sent to global GPU memory. These additional calculations and memory transfers form another overhead, the influence of which we will assess in computational experiments.

We will test two versions of the numerical integration algorithm related to storing blocks of $\bfA^e$ in registers, the first, denoted by {\em REG\_JAC} with assumption that Jacobian matrices are computed on GPU, the second denoted by {\em REG\_NOJAC} assuming that the values of Jacobian matrices are sent to GPU and further read by threads. Both versions are depicted as Algorithm \ref{num_int_GPU_reg}. Whenever the operation of reading is specified without further details, it means a coalasced read from global to shared memory.

\begin{algorithm}
\caption{GPU kernel for numerical integration algorithm with registers used for storing updated entries to element stiffness matrices (details of single updates, the same as for Algorithm \ref{num_int_real}, are omitted here) }
\label{num_int_GPU_reg}
\begin{algorithmic}[1]
\STATE get thread ID within work-group, $thread\_local\_id$ and work-group ID, $group\_id$
\FOR{$ielem=1$ \TO $nr\_elems\_per\_work\_group$}
\STATE read problem dependent coefficients, $\bfc$
\STATE read geometrical data on element
\STATE possibly read ''old'' element degrees of freedom from previous iterations/time steps
\FOR{$ipart=1$ \TO {\em nr\_parts\_of\_stiff\_mat}}
\STATE initialize registers storing entries of element stiffness matrix
\FOR{$i_Q=1$ \TO $N_Q$} 
\STATE read quadrature data, $\bfxi^Q$ and $\bfw^Q$ // {\em (REG\_JAC only)}
\STATE read values of shape functions and their derivatives with respect to local element coordinates, $\bfphi^Q[i_Q]$
\STATE read Jacobian terms {\em (REG\_NOJAC)} or calculate Jacobian matrix, its determinant ({\bf det}) and its inverse {\em (REG\_JAC)}
\STATE calculate derivatives of shape functions with respect to physical coordinates, $\bfpsi^Q[i_Q]$
\STATE based on $\bfc[i_Q]$ and ''old'' degrees of freedom calculate coefficients at quadrature point, $\bfc^Q[i_Q]$
\STATE update a block of entries of $\bfA^e$ associated with a pair $i_{DOF}, j_{DOF}$,
with $i_{DOF}$ and $j_{DOF}$ calculated based on $thread\_local\_id$ and $group\_id$
\ENDFOR
\STATE write the values of registers to  $\bfA^e$ stored in global memory
\ENDFOR
\ENDFOR
\end{algorithmic}
\end{algorithm}

The second strategy for reducing single thread region overhead, mentioned already as an option for parallelization, is to allow threads to compute more blocks for a single quadrature point, by storing them not in registers but in shared memory. This has the advantage of allowing for more explicit management of resources, since in OpenCL the size of shared memory can be queried by the host code. 

In the first strategy, the shared memory may be not fully utilized (it contains problem dependent coefficients, element geometry data and shape functions with derivatives for a single integration point). In the second approach the size of the part of shared memory devoted to store the part of stiffness matrix processed by a single work-group of threads can be explicitly computed, assuming the full utilization of shared memory and taking into account the intended number of active work-groups for a single compute unit. 
As the drawback of this approach the use of shared memory, that may be slower than registers (even assuming properly arranged accesses by all threads) has to be considered. 

The two versions of the algorithm that use shared memory for storing computed parts of element stiffness matrices, {\em SHM\_JAC} with Jacobian terms computed on device and {\em SHM\_NOJAC} with Jacobian matrices sent to GPU global memory are presented as Algorithm \ref{num_int_GPU_shm}. It should be noted, that because of the large size of the element stiffness matrix and usually the small size of the compute unit shared memory, even here, the loop over parts of the stiffness matrix has to be performed.

\begin{algorithm}
\caption{GPU kernel for numerical integration algorithm with shared memory used for storing updated entries to element stiffness matrices (details of single updates, the same as for Algorithm \ref{num_int_real}, are omitted here) }
\label{num_int_GPU_shm}
\begin{algorithmic}[1]
\STATE get thread ID within work-group, $thread\_local\_id$, and work-group ID, $group\_id$
\FOR{$ielem=1$ \TO $nr\_elems\_per\_work\_group$}
\STATE read problem dependent coefficients, $\bfc$
\STATE read geometrical data on element
\STATE possibly read ''old'' element degrees of freedom from previous iterations/time steps
\FOR{$ipart=1$ \TO {\em nr\_parts\_of\_stiff\_mat}}
\STATE initialize part of element stiffness matrix in shared memory
\FOR{$i_Q=1$ \TO $N_Q$} 
\STATE read quadrature data, $\bfxi^Q$ and $\bfw^Q$ // {\em (SHM\_JAC only)}
\STATE read values of shape functions and their derivatives with respect to local element coordinates, $\bfphi^Q[i_Q]$
\STATE read Jacobian terms {\em (SHM\_NOJAC)} or calculate Jacobian matrix, its determinant ({\bf det}) and its inverse {\em (SHM\_JAC)}
\STATE calculate derivatives of shape functions with respect to physical coordinates, $\bfpsi^Q[i_Q]$
\STATE based on $\bfc[i_Q]$ and ''old'' degrees of freedom calculate coefficients at quadrature point, $\bfc^Q[i_Q]$
\FOR{$iblock$ \TO $nr\_blocks\_per\_thread\_within\_part\_of\_stiff\_mat$}
\STATE update a block of entries of $\bfA^e$ associated with a pair $i_{DOF}, j_{DOF}$,
with $i_{DOF}$ and $j_{DOF}$ calculated based on $thread\_local\_id$ and $group\_id$
\ENDFOR
\ENDFOR
\STATE write the entries in the part of stiffness matrix stored in shared memory to  $\bfA^e$ stored in global memory
\ENDFOR
\ENDFOR
\end{algorithmic}
\end{algorithm}

In Algorithms \ref{num_int_GPU_reg} and \ref{num_int_GPU_shm} not all threads are engaged in all kernel operations. In order not to waist the global memory, that may be used by other kernels operating on computed stiffness matrices, there is no padding in global arrays used by kernels.
Since the global data is always accessed in the kernels in a coalesced way, the number of threads performing global read/write operations is specified exactly in the kernels, with some threads remaining idle. 

\begin{figure}
\centering
\includegraphics[width=0.95\hsize]{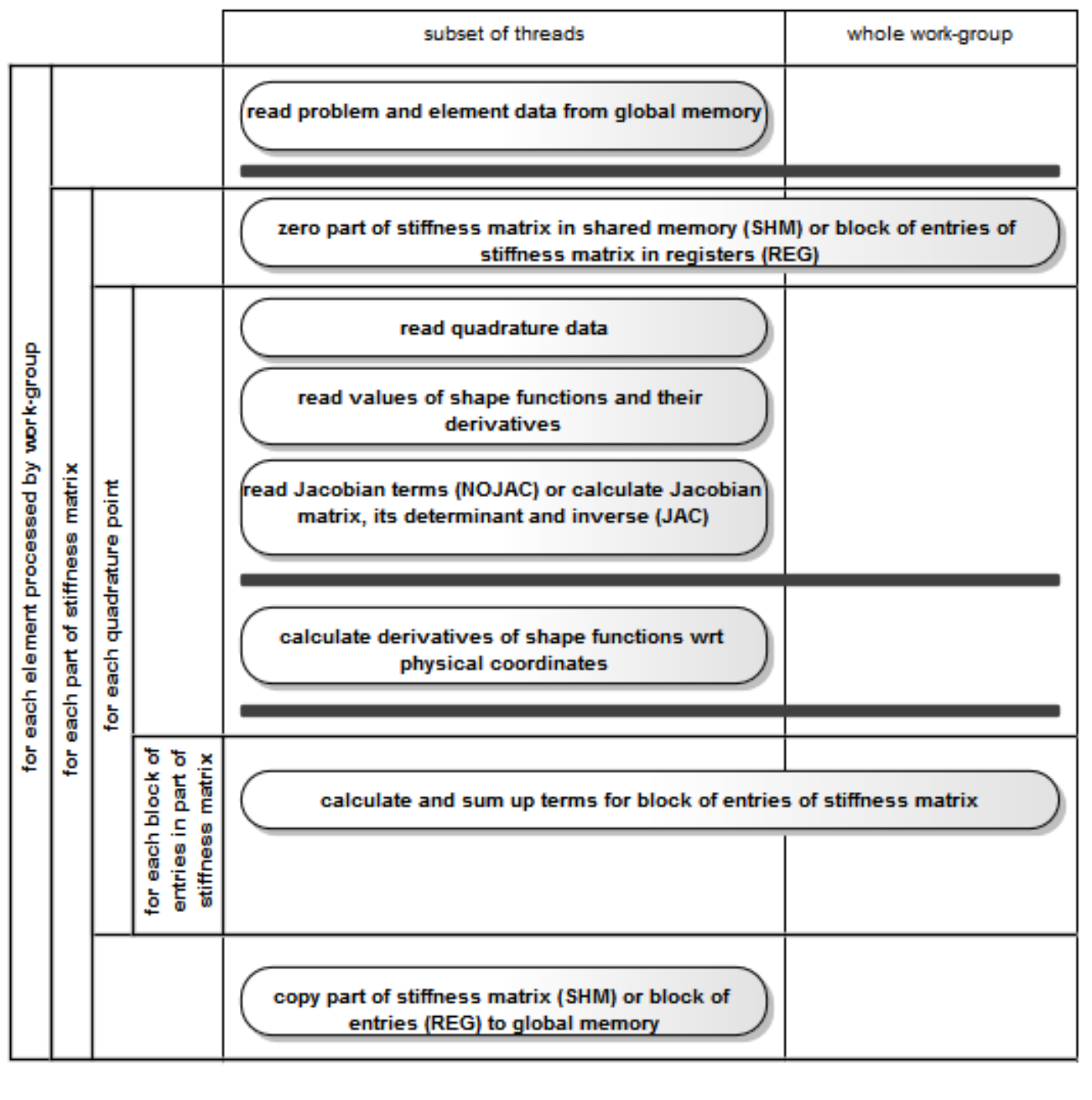}
\caption{Graphical illustration of variants of numerical integration algorithm for GPUs. Work distribution among threads is represented by the width of activity boxes, horizontal bars denote barriers.}
\label{kernel_flow}
\end{figure}

Figure \ref{kernel_flow} presents the kernel workflow with indicated synchronization barriers (using horizontal bars) and the division of work among threads. Different numbers of threads perform different access operations, hence the term ''subset of threads'' denotes different numbers of threads for each read/write operation.

\subsection{Host code design}
Running the kernels described above requires the existence of a CPU code that manages,  on the host side, the execution of GPU functions. We present the host code in the context of large scale calculations, where CPU code may represent one of processes in distributed calculations and the number of elements for single CPU manager thread is large (for higher orders of approximation large may mean tens of thousands, for low order approximations the number may be in the range of millions).

In order to prepare for numerical integration, the host code classifies all assigned to it elements from the mesh or submesh (in distributed calculations) into sets of elements of the same order of approximation. When assembly procedure requires this, each set can be further divided into subsets of elements that do not contribute to the same entries in the created global matrices and vectors (e.g. by some colouring algorithm). Further procedures that we describe concern one of such subsets.

The host code queries the platform on which calculations are performed about the properties of the GPU device. These include: 
\begin{itemize}
\item
the size of GPU global memory (and the maximal size for a single allocated array)
\item
the size of shared (local in OpenCL nomenclature) compute unit memory
\item
the size of constant memory
\item
the maximal number of threads in a work-group i.e. the maximal work-group size 
\item
the maximal number of computing units
\item
the maximal number of threads for kernel
\end{itemize}

After that the CPU code calculates the parameters send as arguments for the GPU kernel. Based on the number of blocks in the element stiffness matrix and the maximal work-group size, the number of threads in a single work-group is computed (as a multiple of 64 as we mentioned in Section \ref{soft_to_hard}). This number, the size of work-group, is maximized, without exceeding other GPU resources and going beyond the number of blocks in $\bfA^e$.

Then, the host code computes the number of parts into which  $\bfA^e$ is divided. For {\em SHM} algorithms it is calculated in such a way that the full available compute unit shared memory is used. This gives the size of the part of stiffness matrix and the number of blocks processed by a single thread in a sequence for the part (it is equal 1 in {\em REG} algorithms). Based on this number and the size of work-group the number of parts of $\bfA^e$ is computed.

In the next preparatory step, the number of elements processed by one kernel invocation is calculated. It is maximized based on the available size for an array storing element stiffness matrices in GPU memory. It is also adjusted to the number of elements in a subset of elements with the same degree of approximation and possibly the same colour.

As the final calculation, based on the number of elements processed by the kernel, the number of elements assigned to a single work-group (associated with the loop in line 2 in Algorithms \ref{num_int_GPU_reg} and \ref{num_int_GPU_shm}) is computed. In order to maximize the number of active work-groups, the number of elements per work-group is minimized. However, still must be large enough so the number of threads does not exceed the limits specified for the GPU device.

After computing parameters for kernel invocation, data forming kernel arguments are created. This include:
\begin{itemize}
\item
the set of calculated or retrieved from finite element data structures parameters of execution -- sent to GPU in an array
\item
quadrature data -- for {\em JAC} versions of algorithms 
\item
geometry related data (nodal coordinates) for elements -- for {\em JAC} variants 
\item
Jacobian terms (determinant of the Jacobian matrix and 9 entries of the inverse Jacobian matrix) for all elements and all quadrature points -- for {\em NOJAC} variants
\item 
array of problem dependent coefficients for elements (possibly combined together in one data structure with geometry data)
\item
the values of all shape functions at all integration points for a reference element (or elements) corresponding to all elements processed by the kernel
\end{itemize}

The prepared data are sent to the GPU memory in several large transfers. The fact of associating as many finite elements as possible with a single kernel invocation should give performance gains, since the overhead associated with data allocation on device and memory transfers is amortized over many elements.

Finally some arrays are allocated in the GPU memory. For {\em SHM} variants  the space for a part of stiffness matrix is allocated in the shared memory. For all variants the workspace for shape functions is allocated in the shared memory and the space for the calculated element stiffness matrices is allocated in the global memory.

The integration kernel is invoked in a usual OpenCL style with the number of threads in a single work-group and the total number of threads for kernel as arguments. 

Computed stiffness matrices either may be copied back from GPU to host memory and transferred to the assembly routine or may be left in GPU memory as an input for some different kernel performing assembly. The first alternative may correspond to situations where there are more elements on the computational node than can fit into GPU global memory, and hence, the GPU performs numerical integration for another portion of elements. For the second alternative, not only assembly can be performed on GPU but the whole solution to the system of linear equations or an update of global vector of unknowns. In such cases computed stiffness matrices are not copied back to the host memory at all.

\section{Computational tests}
\label{computational_tests}

\subsection{Model problem}
As a model problem for numerical examples we consider a 3D linear elastostatics. 
The differential  formulation  of linear elastostatics
for the vector of displacements $u$ = ${u_1, ..., u_d}$, $d=3$ is:
\be
\frac{\partial}{\partial x_j} \left( c_{ijkl} 
\frac{\partial u_{k}}{\partial x_l} \right) = 0, \hspace{2cm} i,j,k,l = 1,..,d
\label{strong_form}
\ee
where $c_{ijkl}$  are in this case  
elasticity coefficients and
summation convention for repeated indices is used. Particular forms
of matrices $\bfc$
depend upon materials considered. The number
of space dimensions $d$ represents, in our case, also the number of equations
in the system and the number of components of the unknown function
$\bfu$.

After discretizing the computational domain $\Omega$ into finite
elements and using a proper basis functions $\hat{\phi}^r$, $r=1,..,N$, 
(the same functions are used for all
components of $\bfu$) the 
approximated displacements $\bfu^h$ are expressed as linear combinations:  
\be
u_k^h =  U_k^r \hat{\phi}^r
\ee
where $U_k^r$ are the discrete unknowns, the coefficients of linear combinations of basis functions.

The resulting final weak statement of the problem that forms a basis for computational procedures
is:
\vspace{2mm}

\noindent
Find a set of coefficients $U_k^r$ such that the following set of
equations is satisfied for all $s=1,..,N$ and $i=1,..,d$:
\be
  U_k^r \int_\Omega c_{ijkl}  
\frac{\partial \hat{\phi}^r}{\partial x_l} 
\frac{\partial \hat{\phi}^s}{\partial x_j} d\Omega
+ BT  = 0
\label{fes}
\ee
where $N \cdot d$ is the total number of scalar unknowns in the system ($N$ is the number of vector unknowns) and 
$BT$ denotes terms corresponding to the boundary
conditions. Numerical integration of these terms is computationally much less
demanding and in our strategy we assume that it is performed by the CPU code.

The domain integral in equation (\ref{fes}) is computed as the sum of element integrals. This corresponds to the fact that the final system of linear equations is assembled from element stiffness matrices, the fact that we adopted as the starting point for the formulation of the numerical integration problem. The analytical expression for a single entry of an element stiffness matrix, in the case of our model problem of linear elasticity is the following:
\be
\sum_{I=1}^{N_Q} c_{ijkl}
\left( \frac{\partial {\phi}^r}{\partial \bfxi}
\frac{\partial \bfxi}{\partial x_l} \right)
\left( \frac{\partial {\phi}^s}{\partial \bfxi}
\frac{\partial \bfxi}{\partial x_j} \right)
\det \bfJ_{\bfT_e} w^Q_I
\label{sum_int}
\ee
where $ \frac{\partial {\phi}^r}{\partial \bfxi}$,$\frac{\partial {\phi}^s}{\partial \bfxi}$ are vectors of derivatives of respective element shape functions with respect to local, reference element coordinates, $w^Q_I$ denotes the weights
associated with particular integration points for a particular quadrature and 
$\bfJ_{\bfT_e}$ denotes the Jacobian matrix of transformation 
${\bfT_e}$ from 
a reference element to the real element $\Omega_e$ and
$\frac{\partial \bfxi}{\partial x_i}$ forms a column of the inverse Jacobian
matrix $\bfJ_{\bfT_e^{-1}}$ corresponding to the inverse transformation ${\bfT_e^{-1}}$). This expression directly corresponds to kernel Algorithms \ref{num_int_GPU_reg} and \ref{num_int_GPU_shm}.

For our model problem of linear elastostatics we assume that PDE coefficients, being material
data, can be different for each element, but are constant over a
single element. In practice, we use only two parameters: Young modulus and Poisson
ratio. The particular form of the array of coefficients implies the operations performed for a single $3\times3$ block of an element stiffness matrix. There are 63 floating point operations necessary in the case of standard updates done by sequential codes. 

The same number of operations for a single block is also performed by the GPU kernels, however, the number of blocks involved is larger, due to the fact that the number of threads in a work-group is a multiple of 64 and is maximized in our kernels. Hence, the number of operations actually performed by hardware is larger for GPUs than for standard processors.

For execution on GPUs apart from floating point operations, there are some operations performed by each thread in the innermost loop of Algorithms  \ref{num_int_GPU_reg} and \ref{num_int_GPU_shm} due to the necessity of computing, for each block, its location in the element stiffness matrix, based on the thread ID and its work-group ID. These calculations involve integer division operations, expensive for GPUs.

\subsection{Hardware set up for testing}
The created kernels have been tested for the full range of considered orders of approximation and several GPUs. For comparison, the same integrations have been performed  on a server with x86 processors.

We present the results for two GPUs on graphics cards: NVIDIA GTX580 and AMD HD5870. The first GPU represents the Fermi architecture \cite{fermi_white_paper}, has 16 streaming multiprocessors and 512 CUDA cores. The GPU memory size is 1.5 GB, with the maximal bandwidth 192.4 GB/s. The number of single precision floating point registers available for threads on a single compute unit is 32786 and the single precision theoretical peak performance is 1581 GFlops. The parameters for the AMD GPU, belonging to the Evergreen family \cite{AMD_APP_guide}, are: 20 compute units, 1600 processing elements, 1 GB global memory, 153.6 GB/s memory bandwidth, 16384 registers and 2.72 TFlops single precision theoretical peak performance. For NVIDIA card we used CUDA 5.0 SDK that provided OpenCL library and for AMD we used APP (version 923.1) environment. Tests for both GPUs were done with Linux operating system and proper driver software provided by NVIDIA and AMD.

The x86 platform consisted of an eight-core Intel XEON E5-2670 processor (2.6 GHz, 332,8 double precision GFLops, 51.2 GB/s maximal memory bandwidth), Linux operating system and Intel C compiler. For comparison with GPUs we have used the standard version of the ModFEM code with double precision calculations and parallelization of numerical integration achieved by parallelizing the loop over elements using OpenMP.

It has to be noted that for comparison we have used an optimized version of CPU code (although for double precision calculations) and we employed a recently (2012) introduced, powerful x86 processor. The two GPU processors are relatively older, by two (NVIDIA) or three (AMD) years, which in the light of the Moore's law and implied speed-ups, 
puts GPUs in less favourable position.

\subsection{Kernel execution characteristics}
Kernel tests were performed for four variants of numerical integration algorithm. We have not introduced specific, hardware oriented optimizations. The same kernels were executed for each order of approximation and each GPU.

For each run the execution begins with retrieving data concerning the platform and the device utilized. These include the size of global memory and the number of compute units (for the values appearing in test runs, see hardware description section) as well as other characteristics, such as the maximal size for memory object allocation (384 MB for NVIDIA and 128 MB for AMD), the size of shared (local in OpenCL terminology) memory (48 kB for NVIDIA, 32 kB for AMD), constant memory size (64 kB for both devices) and the maximal number of threads (work-units) in a work-group (1024 for NVIDIA and 256 for AMD).

Based on the retrieved data, parameters of kernel execution are computed. These parameters are presented in Tables \ref{tab_3} and \ref{tab_4} for the considered orders of approximation $p$ from 2 to 7 and the two GPUs. The only important parameter not included in the table is the number of threads in each work-group. Since we aim at maximizing the number of threads running on a GPU, we use 256 threads per work-group for AMD and 512 threads per work-group for NVIDIA (except the case of $p=2$ where we use 192 threads because of small size of stiffness matrices). The fact that we do not use maximal number of threads per work-group for NVIDIA, results from the large register consumption of our kernels (in the range of 30-37 for AMD and 53-63 for NVIDIA), due to performing several nested loops and complex calculations for a single block of entries of $\bfA^e$.

\begin{table}
\begin{center}
{\footnotesize
\begin{tabular}{|c|c|c|c|c|c|c|}
\hline
\hline
& p=2 & p=3 & p=4 & p=5 & p=6 & p=7 \\
\hline
The number of parts&&&&&& \\
of stiffness matrix -- {\em REG}
& 2 & 4 & 11 & 32 & 76 & 162 \\
\hline
The number of parts&&&&&& \\
of stiffness matrix -- {\em SHM}
& 1 & 2 & 6 & 16 & 38 & 81 \\
\hline
The number of blocks &&&&&& \\
per thread ({\em SHM} only)
&2 & 2 & 2 & 2 & 2 & 2 \\
\hline
The number of elements&&&&&& \\
per kernel
&29056	&5376	&1904	&672	&224	&112 \\
\hline
The number of elements &&&&&&\\
per work-group
&227	&48	&17	&6	&2	&1\\
\hline
The size of input data &&&&&& \\
 for kernel - {\em JAC} [MB]
&8.87	&1.67	&0.67	&0.50	&0.76	&1.52\\
\hline
The size of input data &&&&&& \\
 for kernel - {\em NOJAC} [MB]
&19.95	&9.85	&5.83	&3.92	&2.15	&1.80 \\
\hline
The size of output data &&&&&& \\
for kernel [MB]
&323.21	&295.31	&367.70	&366.28	&295.44	&318.94 \\
\hline
\hline
\end{tabular}
}
\end{center}
\caption{The characteristics of GPU calculations for numerical integration of 3D element stiffness matrices for different orders of approximation $p$ and NVIDIA GeForce GTX580 GPU}
\label{tab_3}
\end{table}

\begin{table}
\begin{center}
{\footnotesize
\begin{tabular}{|c|c|c|c|c|c|c|}
\hline
\hline
& p=2 & p=3 & p=4 & p=5 & p=6 & p=7 \\
\hline
The number of parts&&&&&& \\
of stiffness matrix -- {\em REG}
&2	&7	&22	&63	&151	&324\\
\hline
The number of parts&&&&&& \\
of stiffness matrix -- {\em SHM}
&1	&3	&8	&21	&51	&108 \\
\hline
The number of blocks &&&&&& \\
per thread ({\em SHM} only)
&2	&3	&3	&3	&3	&3 \\
\hline
The number of elements&&&&&& \\
per kernel
&11360	&2240	&640	&160	&80	&40 \\
\hline
The number of elements &&&&&&\\
per work-group
&71	&14	&4	&1	&1	&1\\
\hline
The size of input data &&&&&& \\
 for kernel - {\em JAC} [MB]
&3.47	&0.71	&0.29	&0.34	&0.72	&1.49\\
\hline
The size of input data &&&&&& \\
 for kernel - {\em NOJAC} [MB]
&7.80	&4.11	&1.98	&0.99	&0.88	&0.88 \\
\hline
The size of output data &&&&&& \\
for kernel [MB]
&126.36	&123.05	&123.60	&87.21	&105.51	&113.91 \\
\hline
\hline
\end{tabular}
}
\end{center}
\caption{The characteristics of GPU calculations for numerical integration of 3D element stiffness matrices for different orders of approximation $p$ and AMD Radeon HD5870 GPU}
\label{tab_4}
\end{table}

\subsection{Performance results}
We present performance results for our numerical integration kernels in terms of execution times and GFlops rates. The results in subsequent Tables \ref{tab_5}, \ref{tab_6} and \ref{tab_7} show for subsequent processors, Xeon E5-2670, GeForce GTX580 and Radeon HD5870, the execution times for different phases of numerical integration procedure and the computed performance of calculations. Several remarks below explain the assumptions on which measurements and performance calculations are based.

\begin{table}
 \begin{center}
 {\footnotesize
 \begin{tabular}{|l|r|r|r|r|r|r|}
 \hline
 \hline
\multicolumn{1}{|c|}{Intel XEON E5-2670}  & p=2 & p=3 & p=4 & p=5 & p=6 & p=7 \\
 \hline
 {Flops performed, $\times 10^{-9}$}
 &0.38	 &4.89	 &28.49	 &150.45	 &560.03	 &1757.75\\
\hline
 {Execution time, [$\mu$sec]}
 &9.25	 &104.4	 &672.5	 &3285	 &16180	 &101309 \\
\hline
 {Performance of calculations}
& 41.06	&46.87	&42.37	&45.79	&34.61	&17.35\\
  \hline
 \hline
 \end{tabular}
 }
 \end{center}
 \caption{Execution times on Intel XEON E5-2670(in microseconds), the number of operations performed and the performance of calculations (in GFlops) for the 3D model problem, different orders of approximation $p$ and a single element  stiffness matrix}
 \label{tab_5}
 \end{table}
 
 \begin{table}
 \begin{center}
 {\footnotesize
 \begin{tabular}{|l|r|r|r|r|r|r|}
 \hline
 \hline
 \multicolumn{7}{|c|}{NVIDIA GeForce GTX580} \\
 \hline
 \hline
 Time for: & p=2 & p=3 & p=4 & p=5 & p=6 & p=7 \\
 \hline
 {input data preparation  - {\em JAC}}
 &0.06	&0.08	&0.1	&0.7	&2.9	&15 \\
\hline
 {input data preparation  - {\em NOJAC}}
 &2.00&	5.81&	7.4&	14.5&	24.7&	46\\
\hline
 {GPU memory initialization}
& 11.40	&60.20	&179.5	&481.8	&1144.0	&2457\\
 \hline
 {transfer of input data - {\em JAC}}
&0.07	&0.11	&0.2	&0.7	&3.6	&11\\ 
\hline
 {transfer of input data - {\em NOJAC}}
&0.40	&1.06	&1.9	&4.4	&11.3	&22 \\
\hline
 {transfer of output data}
 & 11.18	 &60.12	 &165.2	 &478.8	 &1137.0	 &2439 \\
\hline
\hline
{\em REG\_JAC} &&&&&& \\
\hline
 {Flops performed, $\times 10^{-6}$} 
 &0.44	 &6.27	 &29.22	 &163.0	 &614.8	 &1981 \\
\hline
 {Execution time for calculations } 
& 2.10	&27.16	&124.0	&679.3	&3010	&9894 \\
\hline
 {Performance of calculations } 
& 208	&230	&235	&240	&204	&200 \\
\hline
{Total execution time } 
 &24.61	&147.6	&469.1&	1641	&5298	&14817 \\
\hline
\hline
{\em REG\_NOJAC} &&&&&& \\
\hline
 {Flops performed, $\times 10^{-6}$}  
 &0.44	&6.30	&29.37	&163.9	&617.9	&1990\\
\hline
 {Execution time for calculations } 
 & 1.21	 &17.80	 &80.07	 &439.9	 &2117	 &7203 \\
\hline
 {Performance of calculations } 
 &366	&354&	366	&372	&291	&276 \\
\hline
{Total execution time } 
 & 25.99	 &149.1	 &442.0 &	1433	 &4458	 &12213 \\
\hline
\hline
{\em SHM\_JAC} &&&&&& \\
\hline
 {Flops performed, $\times 10^{-6}$} 
 &0.43	&6.23	&31.42	&158.9	&590.5	&1868\\
\hline
 {Execution time for calculations } 
 &1.91	&23.06	&114.2	&477.4	&2318	&7357 \\
\hline
 {Performance of calculations } 
 & 228	 &270	 &274	 &274	 &254 &	253 \\
\hline
{Total execution time } 
& 24.41	&143.5	&459.4	&1439	&4605	&12280 \\
\hline
\hline
{\em SHM\_NOJAC} &&&&&& \\
\hline
 {Flops performed, $\times 10^{-6}$} 
 & 0.44	& 6.25	& 33.47	& 159.3	& 592.0	& 1873 \\
\hline
 {Execution time for calculations } 
 &1.30	&19.05	&94.31	&490.9	&1932	&6199 \\
\hline
 {Performance of calculations } 
  &338 &	328 &	334	 &324 &	306	 &302 \\
\hline
{Total execution time } 
 &26.07	 &150.4	 &456.2	 &1484	 &4273	 &11208 \\
  \hline
 \hline
 \end{tabular}
 }
 \end{center}
 \caption{Execution times for different stages of numerical integration on NVIDIA GeForce GTX580 GPU (in microseconds), the number of operations performed and the performance of calculations (in GFlops) for the 3D model problem, different GPU integration algorithms, different orders of approximation $p$ and a single element stiffness matrix}
 \label{tab_6}
 \end{table}
 
\begin{table}
 \begin{center}
 {\footnotesize
 \begin{tabular}{|l|r|r|r|r|r|r|}
 \hline
 \hline
 \multicolumn{7}{|c|}{AMD Radeon HD5870} \\
 \hline
 \hline
 Time for: & p=2 & p=3 & p=4 & p=5 & p=6 & p=7 \\
 \hline
 {input data preparation  - {\em JAC}} 
 &0.03	 &0.04	 &0.2	 &1.58	 &4.8	 &22\\
\hline
 {input data preparation  - {\em NOJAC}} 
& 1.20	&3.21	&5.4	&11.18	&17.1	&52\\
\hline
 {GPU memory initialization} 
&7.50	&37.12	&131.5	&390.4	&902.9	&1995\\
 \hline
 {transfer of input data - {\em JAC}} 
& 0.10	&0.41	&1.7	&6.2	&13.6	&31\\
\hline
 {transfer of input data - {\em NOJAC}} 
&0.58	&1.74	&4.1	&16.9	&27.9	&60 \\
\hline
 {transfer of output data} 
&2.41	&12.19	&42.5	&162.8	&318.4	&639\\
\hline
\hline
{\em REG\_JAC} &&&&&& \\
\hline
 {Flops performed, $\times 10^{-6}$} 
 &0.43	 &5.56	 &30.05	 &168.6	 &659.0	 &2206 \\
\hline
 {Execution time for calculations } 
&3.13	&53.73	&325.3	&1779	&9496	&34568 \\
\hline
 {Performance of calculations } 
&139	&103	&92	&94	&69	&63\\
\hline
{Total execution time } 
 &10.77	 &91.32	 &458.9	 &2177	 &10418	 &36618 \\
\hline
\hline
{\em REG\_NOJAC} &&&&&& \\
\hline
 {Flops performed, $\times 10^{-6}$}  
&0.44	&5.62	&30.36	&170.2	&665.1	&2226 \\
\hline
 {Execution time for calculations } 
 &4.06	&34.61	&190.8	&985	&4684	&17094 \\
\hline
 {Performance of calculations } 
 &109	 &162	 &159	 &172	 &141	 &130 \\
\hline
{Total execution time } 
 &13.35	 &76.68	 &332.0	 &1404	 &5632 &19202 \\
\hline
\hline
{\em SHM\_JAC} &&&&&& \\
\hline
 {Flops performed, $\times 10^{-6}$} 
&0.43&	7.02	&31.57	&157.8	&602.5&	1906\\
\hline
 {Execution time for calculations } 
& 3.90	&63.74	&297.5	&1477	&6428	&21359 \\
\hline
 {Performance of calculations } 
& 111	&110	&106	&106	&93	&89 \\
\hline
{Total execution time } 
&11.54	&101.3	&431.0	&1876	&7350	&23408\\
\hline
\hline
{\em SHM\_NOJAC} &&&&&& \\
\hline
 {Flops performed, $\times 10^{-6}$} 
 &0.44	 &7.05	 &24.59	 &158.3	 &604.6	 &1912\\
\hline
 {Execution time for calculations } 
& 5.33	&67.90	&305.9	&1519	&6152	&20742 \\
\hline
 {Performance of calculations } 
 &82	&103	&103	&104	&98	&92\\
\hline
{Total execution time } 
& 14.62	&109.9	&447.1	&1937	&7101	&22850 \\
  \hline
 \hline
 \end{tabular}
 }
 \end{center}
 \caption{Execution times for different stages of numerical integration on AMD Radeon HD5870 GPU (in microseconds), the number of operations performed and the performance of calculations (in GFlops) for the 3D model problem, different GPU integration algorithms, different orders of approximation $p$ and a single element stiffness matrix}
 \label{tab_7}
 \end{table}
 
 \begin{itemize}
\item
Each presented table contains timing results for subsequent stages of the whole procedure of numerical integration. This reflects the possibility of arranging the stages in different ways. The initial stage is important when the first group of elements is sent for integration on GPU. Then, for next groups it is possible not to allocate the GPU memory and send only input data for processed elements. Also the stage of copying back the computed stiffness matrices can be eliminated in certain solution strategies, as we have mentioned already.   
\item
The timing results are presented for a single finite element of a given order. These results were obtained by running the kernels for a large number of elements (specified for our test runs in Tables \ref{tab_3} and \ref{tab_4}) and dividing the times of performing different stages of execution by the number of elements. It should be made clear that the results depend on the total number of elements, since the times for card initialization and data transfers to and from GPU memory does not scale linearly with the number of elements.
\item
The nonlinear dependence mentioned above can appear for low numbers of elements. For sufficiently high numbers, the linear dependence should emerge and the results presented in Tables \ref{tab_5}--\ref{tab_7} should be applicable to different problem sizes by simple scaling. The linear scaling should be observed especially for higher orders of approximation where GPU calculations, that scales linearly, take more time than card initialization and data transfers.
The fact of linear scaling of the numerical integration algorithm with respect to the number of elements for sufficiently high numbers of elements has been confirmed experimentally in several studies \cite{ppam_09_gpu,Markall_2013,Cecka_2011}.
\item
As it was mentioned already, for CPU cores and GPU compute units we parallelize the loop over elements. If the processor resources are sufficient, than, thanks to the embarrassingly parallel character of the loop over elements in the numerical integrations algorithm, the performance of calculations should scale linearly (even in the strong sense) with the number of cores and compute units. We observed such relations for different GPUs with the NVIDIA Tesla architecture when performing computational experiments reported in \cite{iccs_10_gpu,imcsit_10}.
\item
It should be noted, that both types of scaling, with respect to the number of elements and with respect to the number of cores/compute units, are much more complex when assembly process is taken into account. In such a case also the dependence of performance results on a particular problem solved (due to the different properties of the global stiffness matrix) can be observed \cite{Knepley_2013,iccs_hp_solver_2013}. In the case of numerical integration alone, the particular problem solved is irrelevant to the performance obtained.
\item
Execution times were always measured on the host side. We introduced synchronisation procedures for OpenCL events in the host code and measured elapsed times for each stage of calculations. The reason for this was to explicitly indicate the overhead associated with the subsequent stages. In recent modifications to GPU programming models it is possible to overlap different stages of execution for different kernel invocations and reduce the overheads \cite{cuda_guide}.
\item
Measuring elapsed time means adopting a user perspective. We do it for particular stages of calculations and also for the total execution time of kernels. The times measured on the host side were always longer than the times reported by execution environment profilers. Hence the performance that we report is the performance observed by the user, which is lower than the actual performance of the hardware. 
\item
Times for GPU initialization were relatively constant for all orders of approximation since they were mainly associated with creation of OpenCL memory objects in GPU memory and the total size of memory objects for each order was similar, due to the assumption that by changing the number of elements per kernel the whole available GPU memory is utilized. However, for each order of approximation the times per element are different, which is implied by the changing size of the stiffness matrix for a single element.
\item
For computing performance in GFlops, only the time for kernel execution (measured on the host side) was taken into account.
\item
To obtain the total execution time, we measured, in an external procedure responsible in our code for global matrix assembly, the time spent in the host code and the integration kernels. 
\end{itemize}



\begin{figure}
\centering
\includegraphics[width=0.45\hsize]{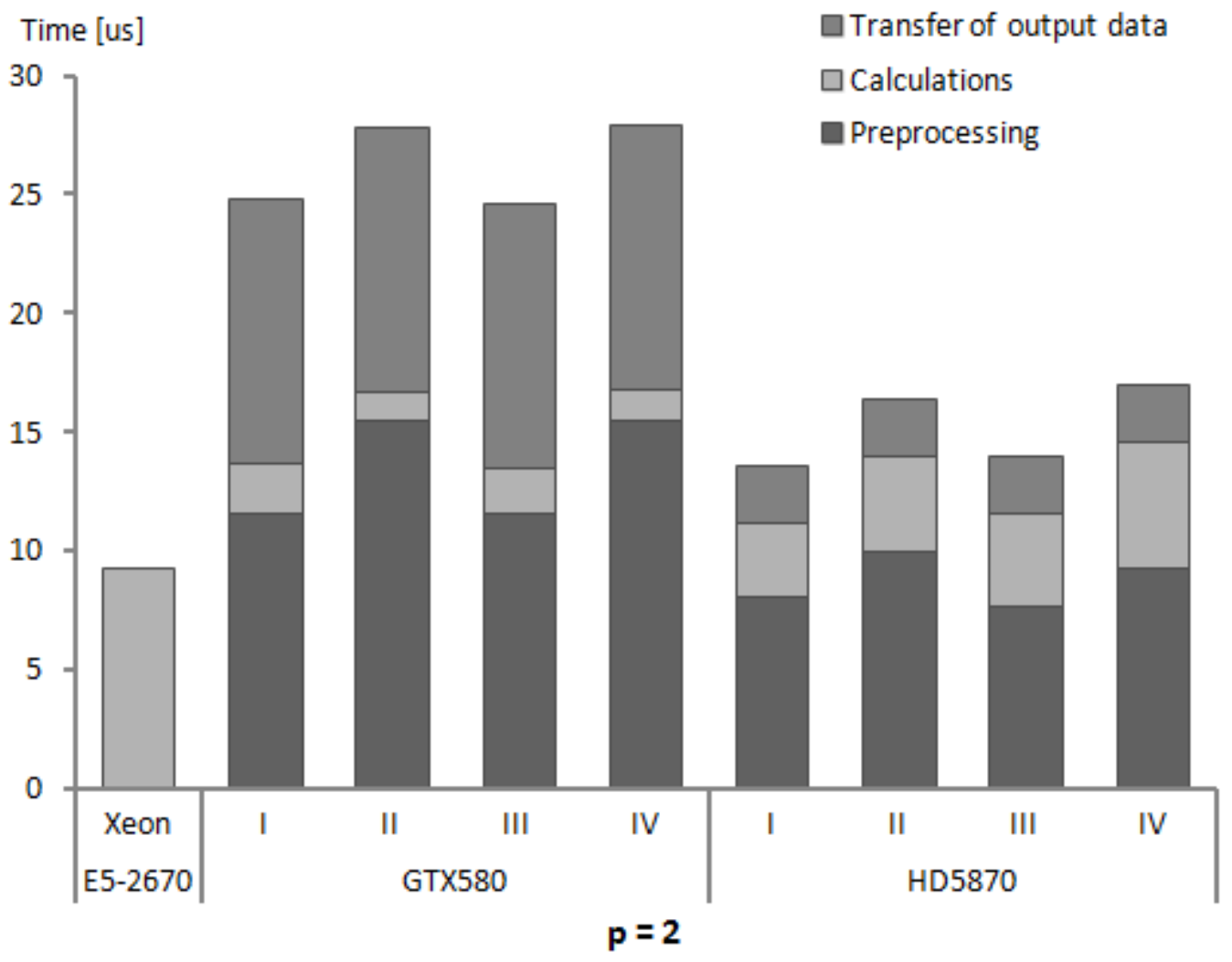} ~ ~ ~
\includegraphics[width=0.45\hsize]{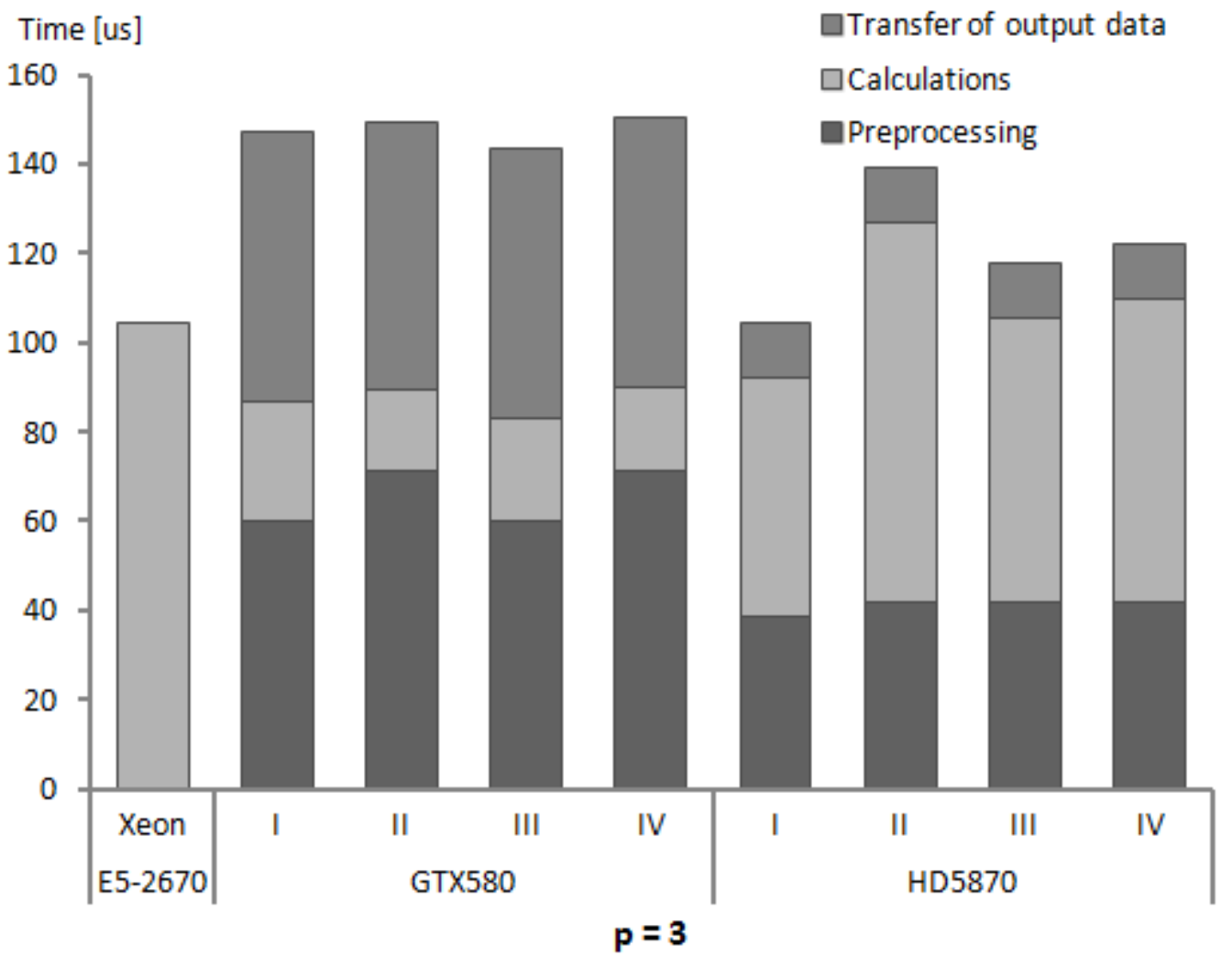} 
\includegraphics[width=0.45\hsize]{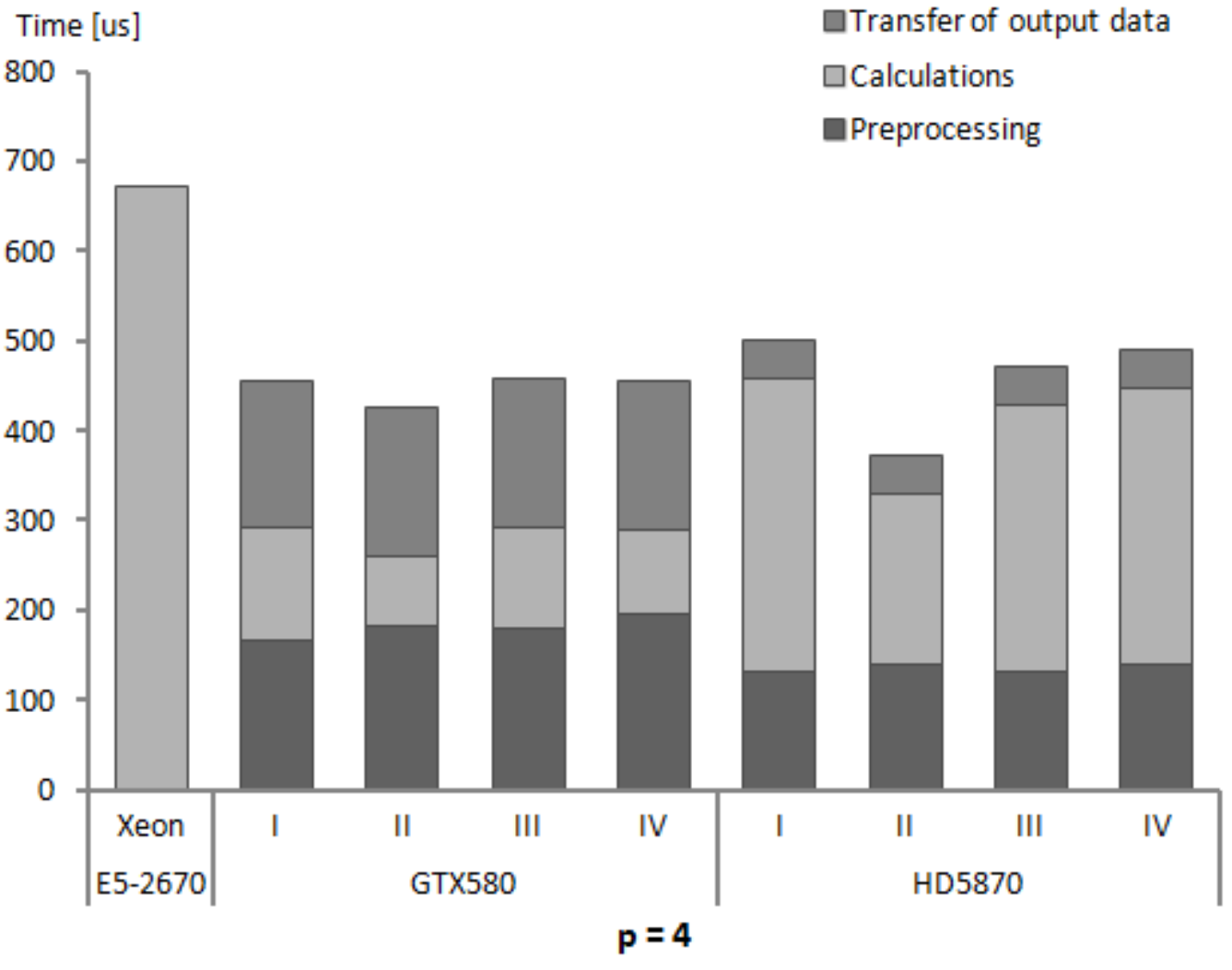} ~ ~ ~
\includegraphics[width=0.45\hsize]{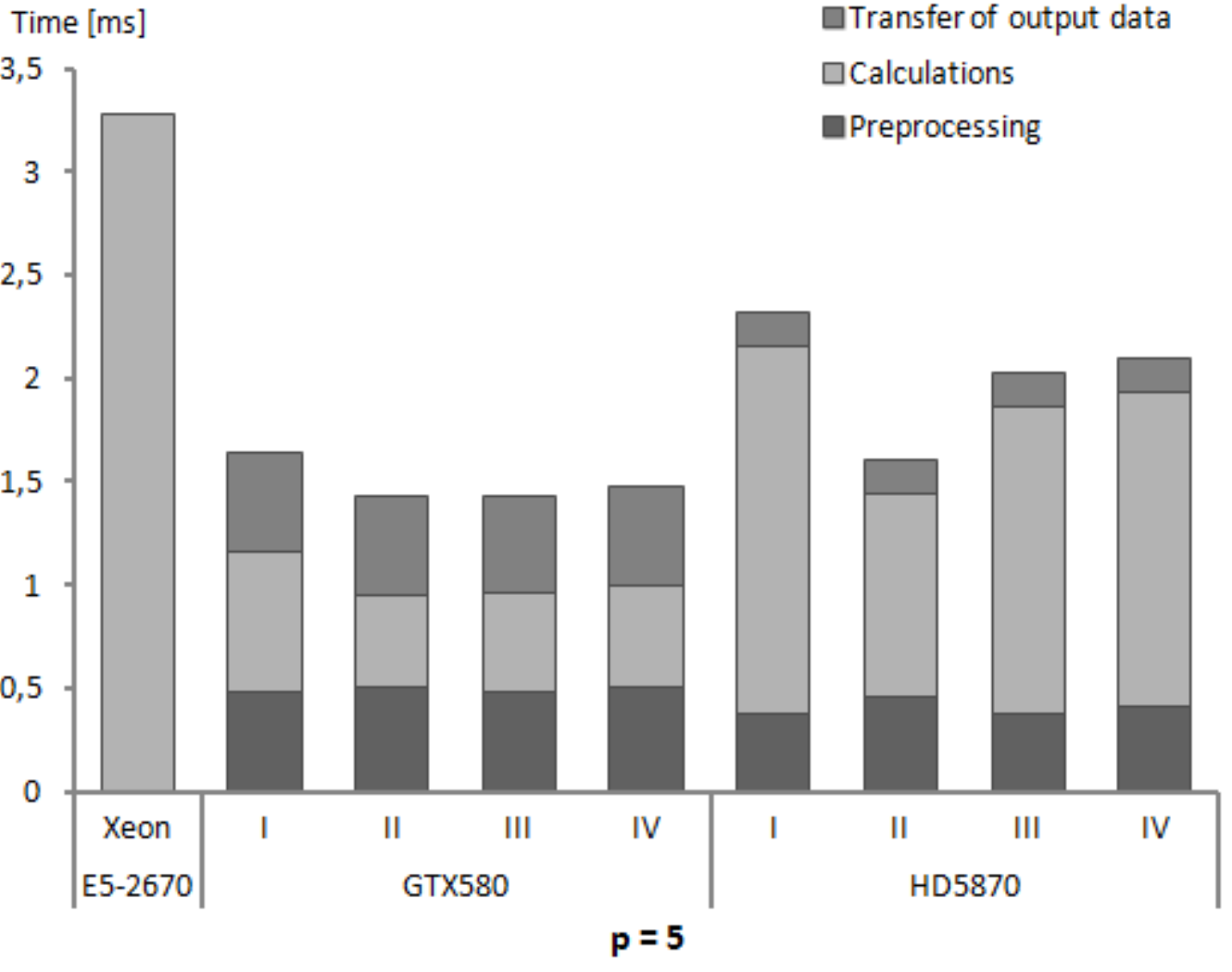}
\includegraphics[width=0.45\hsize]{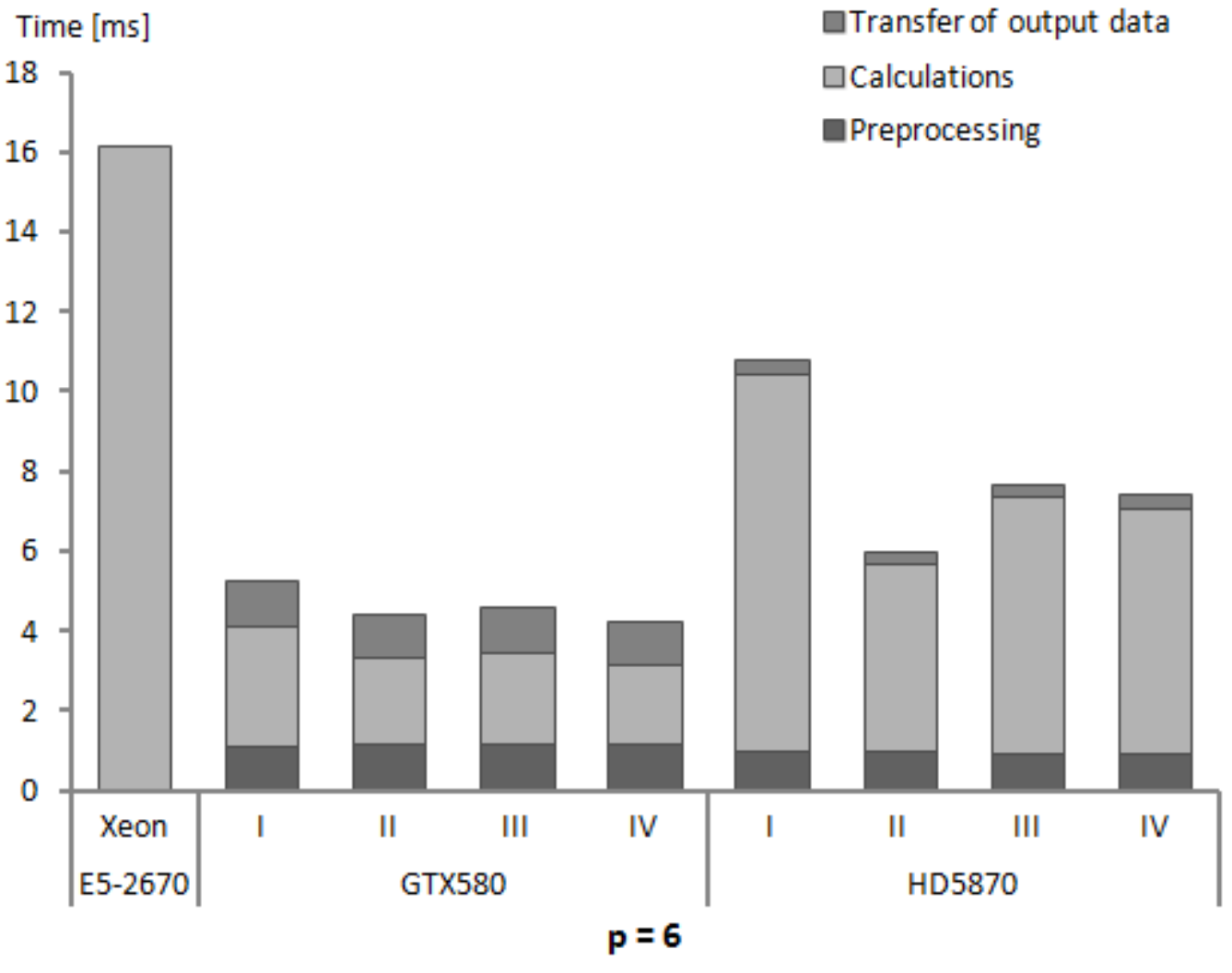} ~ ~ ~
\includegraphics[width=0.45\hsize]{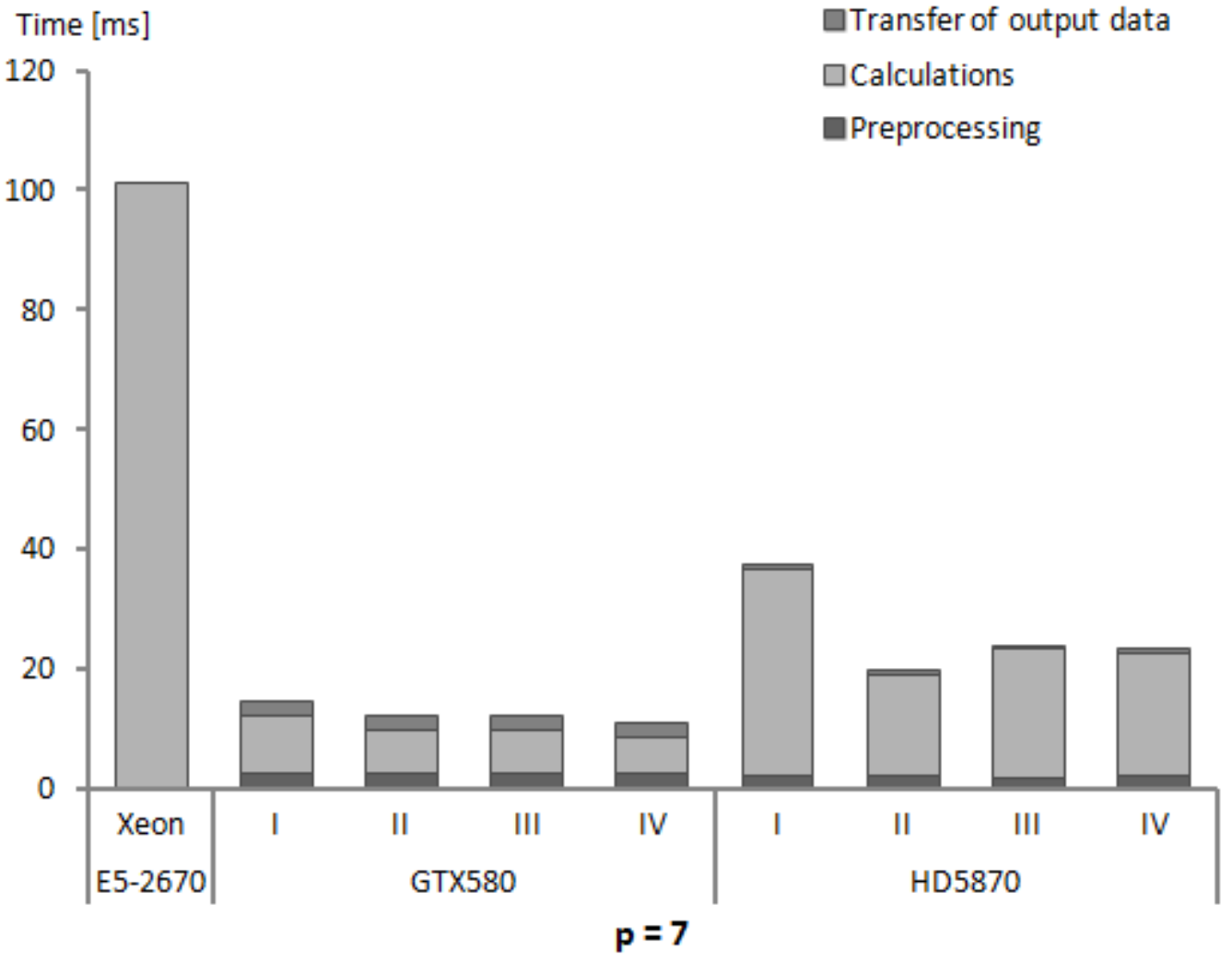}
\caption{Execution times for numerical integration procedures for one element and different processors: Intel Xeon E5-2670, NVIDIA GeForce GTX580 and AMD Radeon HD5870. Different variants of GPU kernels are denoted by: I - {\em REG\_JAC}, II - {\em REG\_NOJAC}, III - {\em SHM\_JAC}, IV - {\em SHM\_NOJAC}. GPU times are split into preprocessing phase (preparation of input data, initialization of GPU memory and transfer of input data for kernels), calculations (kernel execution) and transfer of output data produced by kernels. Subsequent figures present times for orders of approximation from 2 to 7.}
\label{times}
\end{figure}

\begin{figure}
\centering
\includegraphics[width=0.45\hsize]{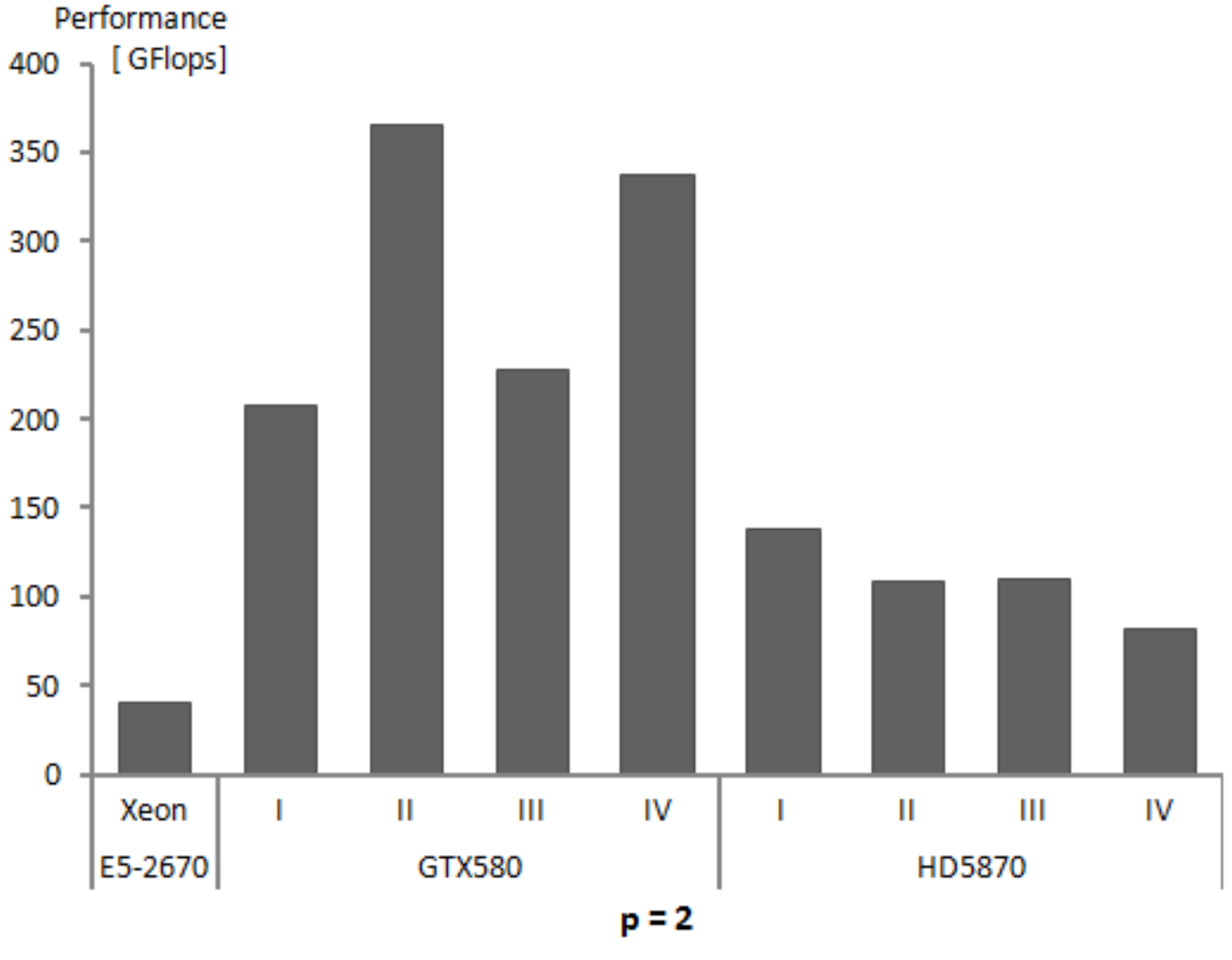}  ~ ~ ~
\includegraphics[width=0.45\hsize]{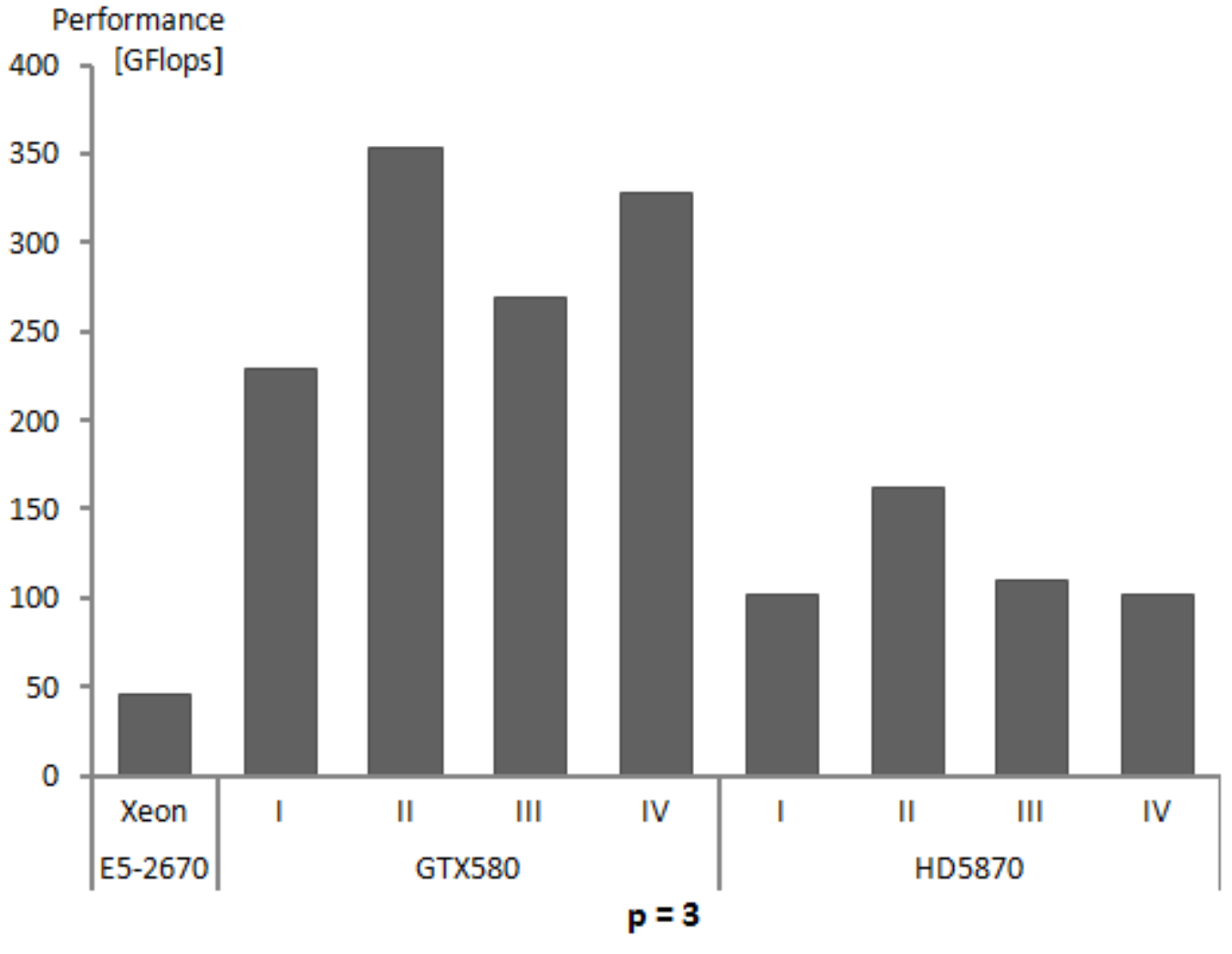}
\includegraphics[width=0.45\hsize]{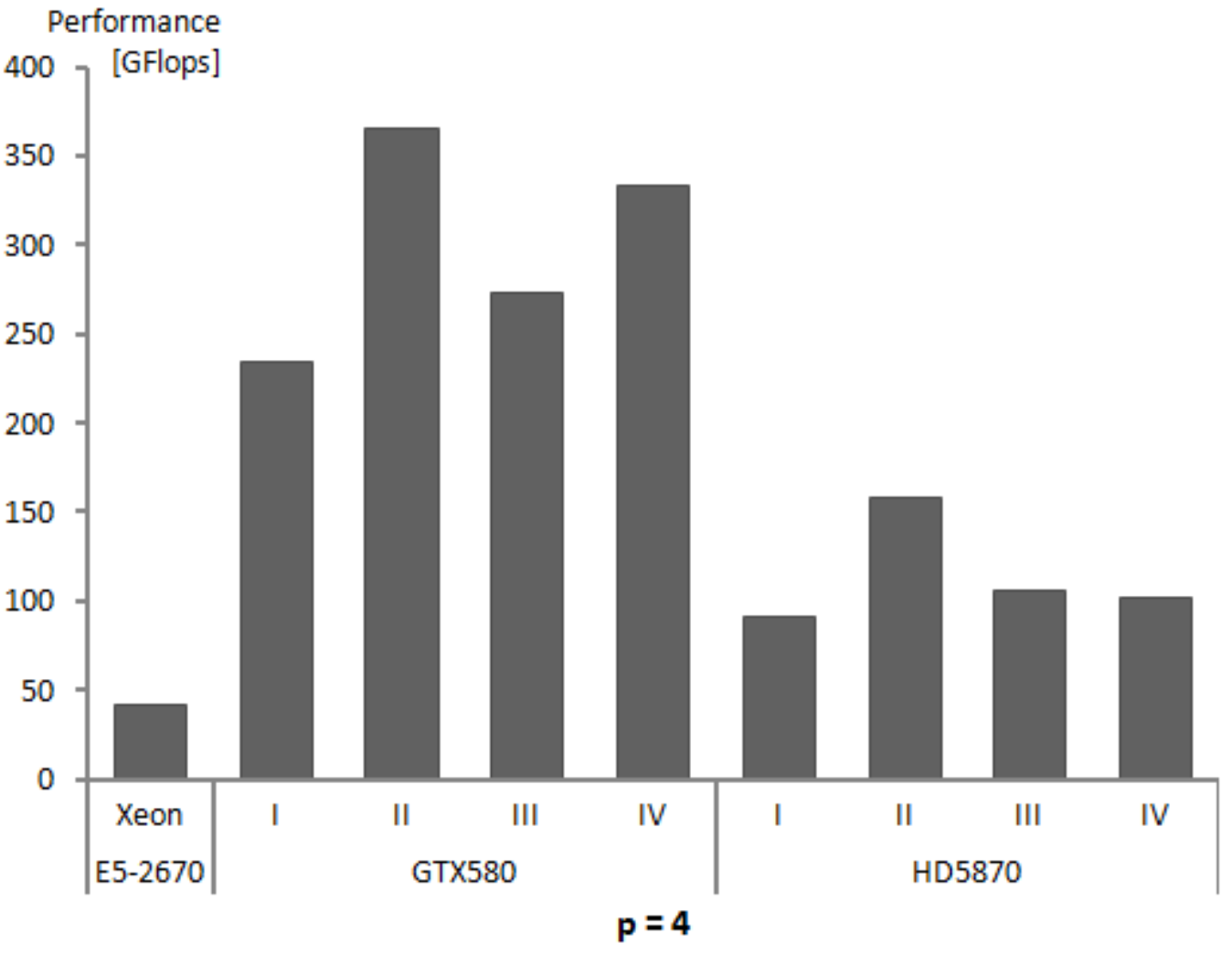}  ~ ~ ~
\includegraphics[width=0.45\hsize]{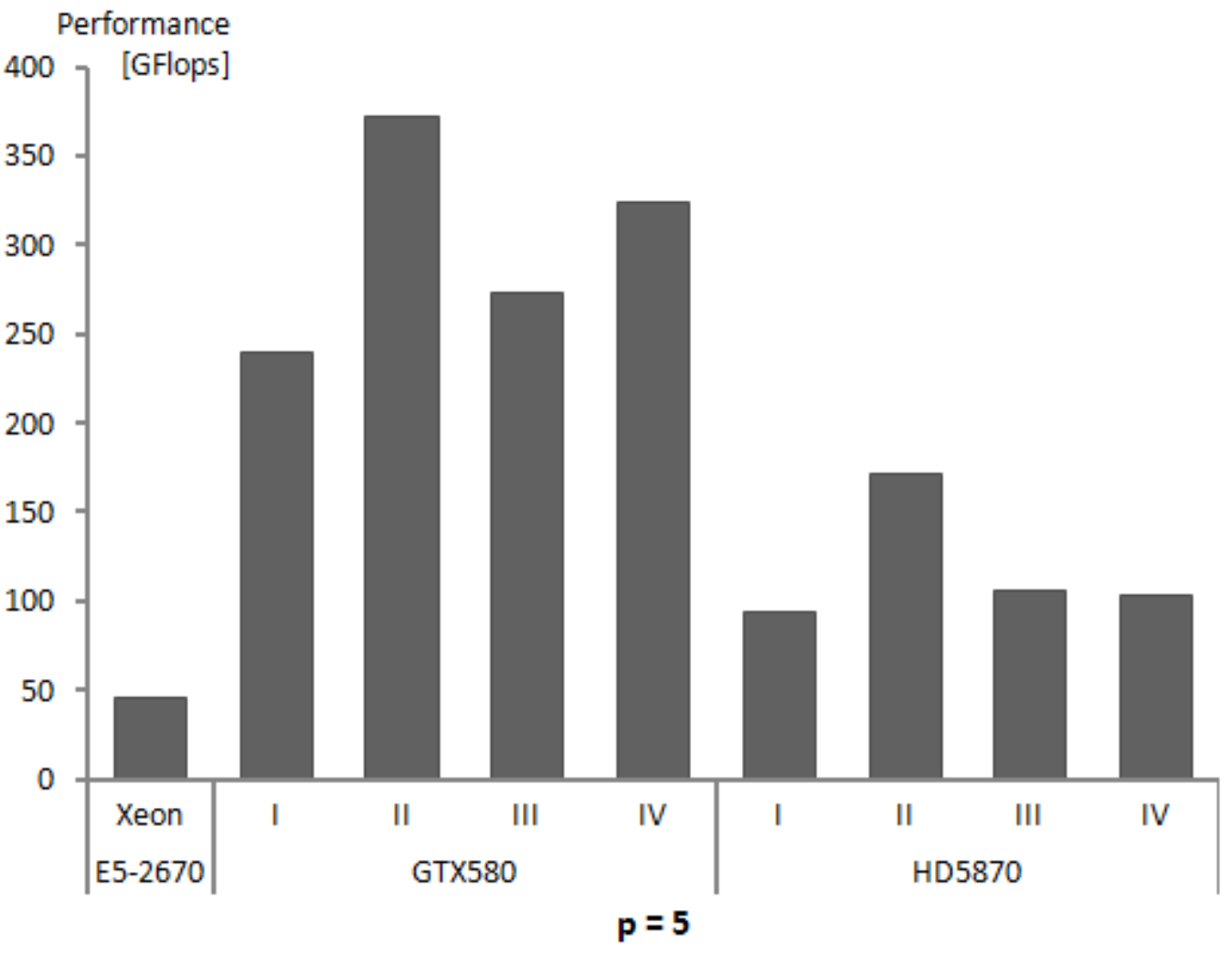}
\includegraphics[width=0.45\hsize]{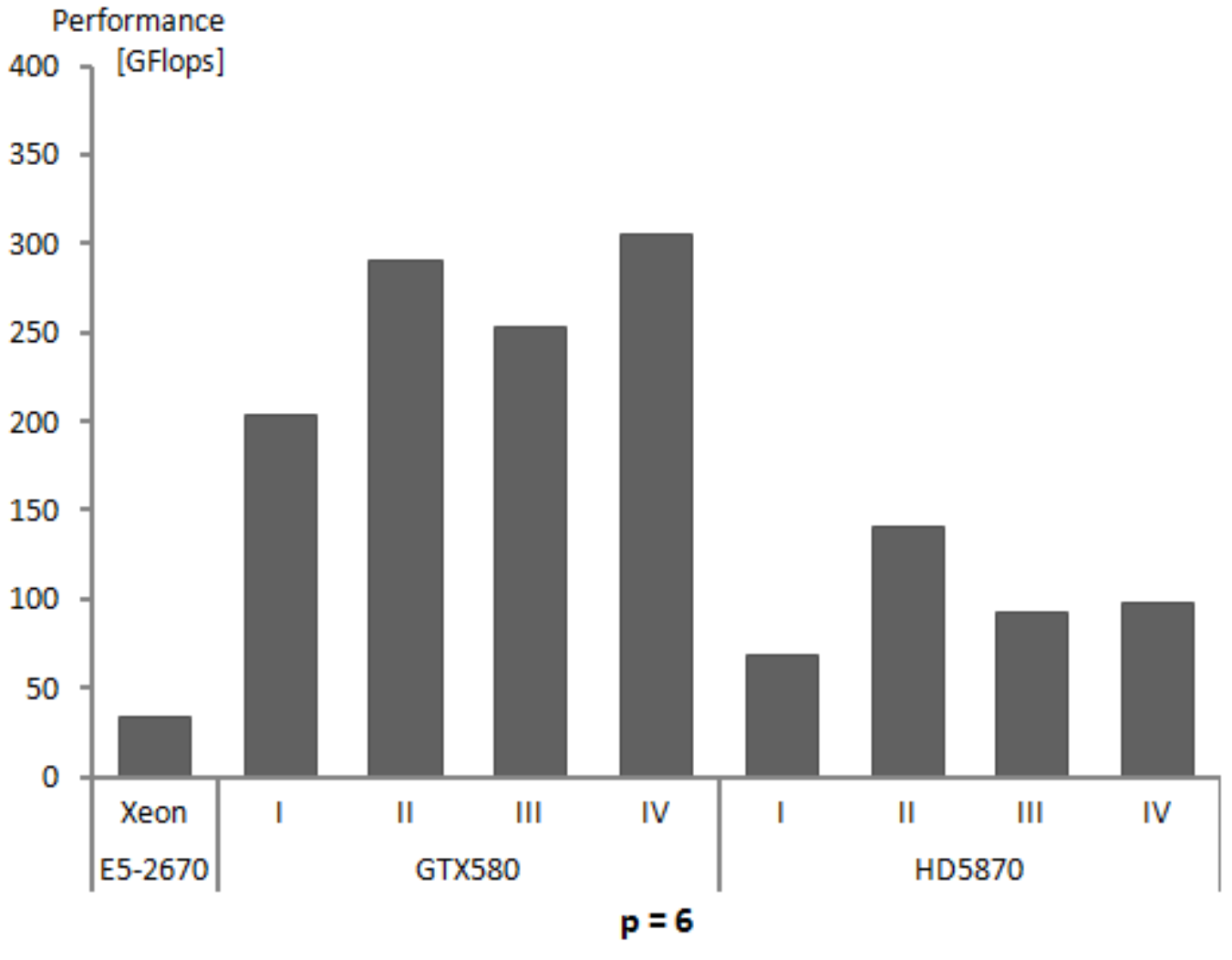}  ~ ~ ~
\includegraphics[width=0.45\hsize]{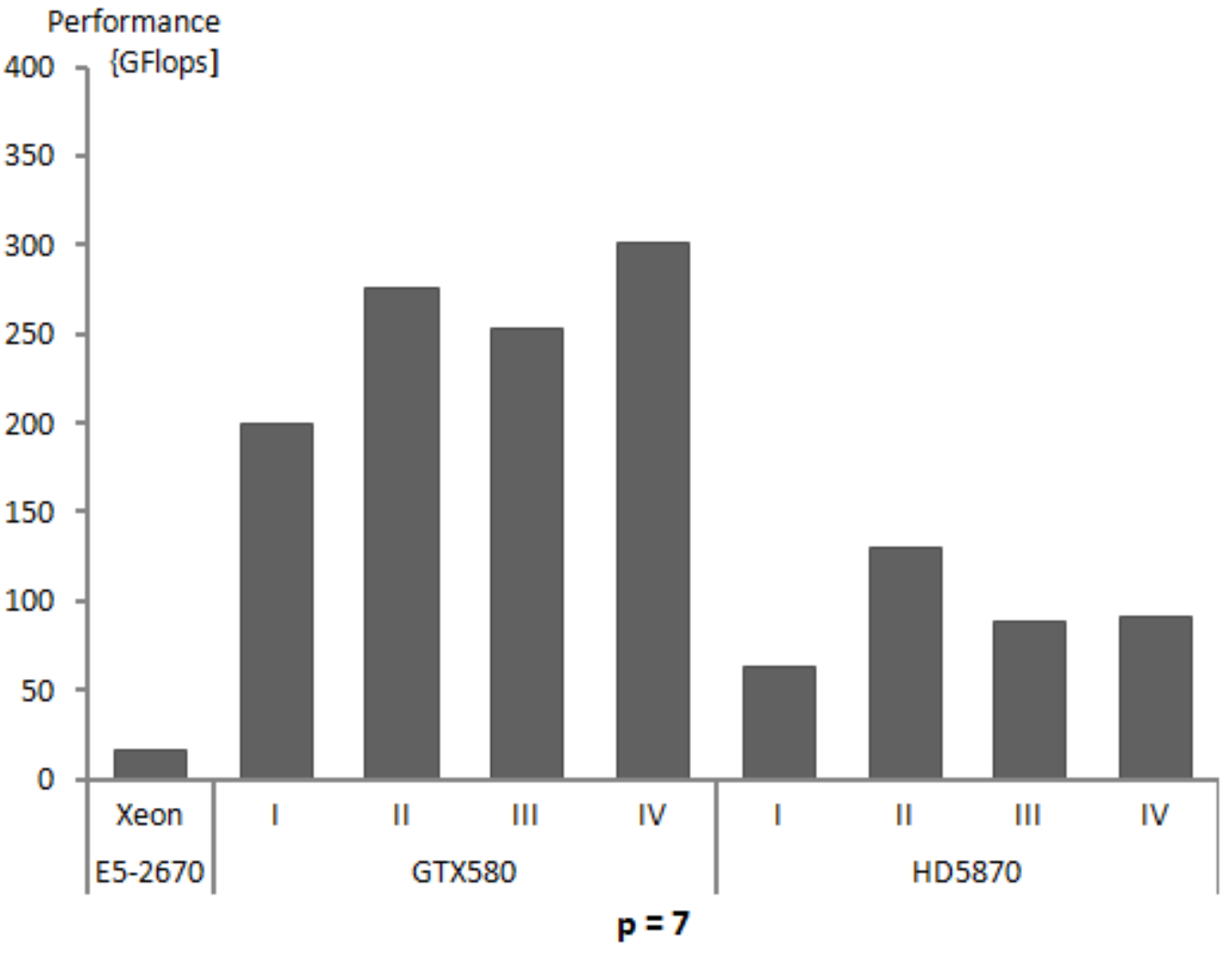}
\caption{Performance of numerical integration calculations for one element and different processors: Intel Xeon E5-2670, NVIDIA GeForce GTX580 and AMD Radeon HD5870.Different variants of GPU kernels are denoted by: I - {\em REG\_JAC}, II - {\em REG\_NOJAC}, III - {\em SHM\_JAC}, IV - {\em SHM\_NOJAC}. Subsequent figures present performance in GFlops for orders of approximation $p$ from 2 to 7.}
\label{performance}
\end{figure}

The main results from Tables \ref{tab_5}, \ref{tab_6} and \ref{tab_7} are illustrated in Figures \ref{times} and \ref{performance}

\subsection{Discussion of results}

There are several observations that can be made based on data in Tables \ref{tab_5}, \ref{tab_6} and \ref{tab_7}:
\begin{description}
\item[Observation 1.]
The algorithm of numerical integration for higher order finite elements has the potential for high performance execution on massively multi-core architectures. The performance obtained, above 200 GFlops, and for some orders of approximation and some variants of the algorithm above 350 GFlops for GTX580 and above 60 GFlops with several results above 150 GFlops for HD5870, can be considered high, especially for NVIDIA where it is in the range of 12\%-23.5\% of the peak single precision floating point performance.
\item[Observation 2.]
The performance of GTX580 GPU is much higher than that of HD5870, despite the fact that Radeon HD5870 theoretical peak performance is almost twice as big as that of GeForce GTX580. This can be explained to certain degree by higher overhead due to driver software for Radeon \cite{Markall_2013} and the fact that basic execution units of HD5870 are vector units, the fact that we do not exploit in our kernels. 
\item[Observation 3.]
For GPUs, the times for global memory initialization and data transfers from the host memory are comparable to the time of kernel execution (for lower values of $p$ being even several times larger). This, once again, here for finite element numerical integration algorithm, confirms the fact that slow PCIe connection is an obstacle on a way to get higher performance for general purpose codes executed on GPUs. For the algorithm of numerical integration, with increasing order of approximation the ratio of operations to memory transfers grows and hence the influence of low PCIe performance diminishes.
\item[Observation 4.]
The times of kernel execution, i.e. the times for numerical integration calculations are always much lower for GPUs than for the reference CPU, however, the total execution times, including memory transfers are better for GPUs only in the case of higher orders of approximation.
\item[Observation 5.]
There is a large difference between the performance of the GPU as perceived by an external user and as exhibited during actual calculations. One reason for this, data transfer times, was already mentioned. The second is the fact, that, in order to ensure proper mapping of calculations onto the hardware, the GPU had to perform more operations than the CPU.
\item[Observation 6.]
There is no single kernel being the best for all situations (different orders of approximation, different hardware), although the {\em REG\_NOJAC} variant is the fastest in most of cases.
Different kernels exhibit similar performance, with the best results obtained for medium orders of approximation.
\item[Observation 7.]
The option of sending to GPUs, not only reference element shape functions, but also Jacobian terms for all real processed elements and all integration points ({\em NOJAC} versions) turned out to be more efficient in most cases than the option of calculating the values on GPUs ({\em JAC} variants).
Despite the fact that {\em NOJAC} versions require much longer times for transferring data to GPU memory than {\em JAC} variants, this time remains much shorter than GPU initialization time and time for transfer of output data.
\item[Observation 8.]
{\em SHM} variants are usually slower than {\em REG} versions, however we believe it is worth having them both, since for different GPU architectures this situation may change.
\end{description}


\section{Conclusions}
\label{conclusions}

Numerical integration for higher order 3D finite element approximations is a complex problem due to the large variation in required processor and memory resources. Thanks to a suitable parametrization of the designed OpenCL code for integration and a proper management of code execution by the created host side procedures, we have obtained the code that is portable across different GPU architectures and different orders of approximation, in the range of moderate degrees of shape function polynomials, from 2 to 7. In the computational experiments reported in the paper we run the GPU kernels and the host managing procedures for all presented cases without changing a single line in the source code. 

We obtained, for NVIDIA GeForce GTX580 and AMD Radeon HD5870 graphics processors, the performance of calculations always above 200 GFlops and 60 GFlops, respectively. This, especially for HD5870, falls short to theoretical peak performance of the hardware, however still for many cases reported, the results can be considered satisfactory as for general purpose complex scientific calculations. In particular, for the case of GTX580 and one of the variants of numerical integration ({\em SHM\_NOJAC}) the performance for all orders of approximation exceeded 300 GFlops (19\% of theoretical peek) and for another variant ({\em REG\_NOJAC}) it reached for three orders ($p$=2, 4 and 5) more than 365 GFlops, that is 23\% of the theoretical maximum.

The main factor limiting the practical advantages of GPU calculations is the slow PCIe connection between the GPU and the host computer. Despite much higher GPU performance of calculations, the overall execution times for GPUs and standard processors are comparable, because of long times devoted by GPUs to memory initialization and data transfers.
 
Taking this into account, it can be concluded that in the context of numerical integration for higher order finite elements, GPUs should not be considered as a replacement for CPUs, with all calculations off-loaded to GPUs. 
In particular, when GPUs are calculating domain
integrals, CPUs can do some other useful work, such as performing integration of boundary terms, integration of separate sets of elements or assembly of already created element stiffness matrices. 
In such scenarios, GPUs can form  valuable accelerators increasing substantially the overall performance of finite element codes.

\subsubsection*{Acknowledgements.}
 The support of this work by the Polish National Science Centre under grant no DEC-2011/01/B/ST6/00674 is gratefully acknowledged.












\end{document}